\documentclass{article}

\usepackage[main, final]{neurips_2026}

\usepackage[utf8]{inputenc}
\usepackage[T1]{fontenc}
\usepackage{amsmath,amssymb,amsfonts}
\usepackage{graphicx}
\usepackage{booktabs}
\usepackage{hyperref}
\usepackage{url}
\usepackage{nicefrac}
\usepackage{microtype}
\usepackage{xcolor}
\usepackage{algorithm}
\usepackage{multirow}
\usepackage{algpseudocode}
\usepackage{placeins}
\usepackage{float}
\usepackage{tikz}
\usetikzlibrary{positioning,arrows.meta,fit}
\usepackage{enumitem}

\title{LeCellModel: Interpretable Density Estimation over the Gene Expression Manifold}

\author{%
    Gil Karin\thanks{Equal contribution.} \quad
    Artemii Bakulin\footnotemark[1] \quad
    Nir Yosef \\
    Weizmann Institute of Science \\
    \texttt{\{gil.karin, artemii.bakulin, nir.yosef\}@weizmann.ac.il}
}

\begin{document}
\maketitle
\begin{abstract}

Representation learning for single-cell RNA-seq has been evaluated by the identities representations capture — cell types, lineages, perturbations — rather than how faithfully they
model the distribution of cell states on the gene expression manifold. As a result, density estimation, rare-state detection, perturbation scoring, and gene-level attribution are
bolted onto frozen embeddings as separate models, disconnected from the geometry of the data. We argue that modeling the distribution of cell states should be the primary objective, and that representation quality follows from it. Building on LeJEPA, whose SIGReg objective enforces a provably optimal isotropic Gaussian latent, we introduce LeCellModel, which reads the encoder's Jacobian log-volume (JEPA-SCORE) under this Gaussian latent as a relative typicality score. Across several datasets, LeCellModel outperforms established methods on the scIB
benchmark for representation quality. Furthermore, we show that LeCellModel enables detection of rare expression states using JEPA-SCORE as a typicality measure, applying it to a
controlled cytokine panel, where it identifies stimulated cells across minority fractions and tracks pathway-response magnitude. We introduce gene-level interpretability of the
typicality score — to our knowledge, the first method for analyzing JEPA-SCORE in feature space. Leveraging this framework, LeCellModel recapitulates the known association of a
profibrotic macrophage program with idiopathic pulmonary fibrosis, and discovers a CD177$^{+}$ activated regulatory T cell subpopulation previously described only in tumor
contexts. Both findings emerge directly from the geometry of the learned embedding, without relying on supervised annotations. The code is available at
  \url{https://github.com/gilk717/lecellmodel_paper}.

\end{abstract}

\section{Introduction}

Single-cell RNA sequencing measures gene expression across thousands of individual cells, producing high-dimensional count data that capture the molecular state of each cell. A central computational task is to learn representations of these states that organize cells into biologically meaningful structure: separating cell types, resolving developmental trajectories, and exposing responses to perturbation. The dominant approach has been to train encoder-decoder models with reconstruction objectives, from variational autoencoders such as scVI~\citep{scVI} and scANVI~\citep{scANVI} to BERT-style foundation models such as Geneformer~\citep{geneformer} and scGPT~\citep{scGPT}. Reconstruction, however, operates entirely in gene space, which encourages the encoder to retain nuisance variation such as batch effects and dropout patterns alongside biological signal. Adversarial batch correction methods like scDREAMER~\citep{scDREAMER} partly offset this, but the adversary works against a reconstruction objective that itself rewards retaining batch-related variation to minimize error. Finally, there are emerging approaches, such as GeneJEPA~\citep{geneJEPA} and CellJEPA~\citep{cellJEPA}, following the Joint-Embedding Predictive Architecture (JEPA) paradigm that forgo the decoder entirely, shaping the embedding space through latent prediction rather than reconstruction. Crucially, all these methods are judged primarily by the identities their representations capture, rather than by how faithfully they describe the distribution of cell states.

A remarkable yet largely unexploited property of JEPAs is that, when the latent space is constrained to an isotropic Gaussian, the encoder provably estimates the density of the
training distribution~\citep{gaussian-embeddings}. We are the first to leverage this property in a practical single-cell setting, introducing LeCellModel: a JEPA-based model for
learning the distribution of cell states. LeCellModel builds on LeJEPA \citep{lejepa-provable}, which unlike previous single-cell JEPAs avoids heuristics such as stop-gradient to
prevent collapse, replacing them with a principled isotropic Gaussian latent enforced via Sketched Isotropic Gaussian Regularization (SIGReg) \citep{lejepa-provable}. In addition, rather than scaling to large transformer architectures, we show that a lightweight MLP encoder is sufficient for single-cell data, keeping the model tractable and the Jacobian
computable. Across several datasets on the scIB benchmark \citep{scib-metrics}, LeCellModel achieves state-of-the-art representation quality, while its typicality score, read
within a cell type, enables practical detection of rare cell states. Moreover, we show that differentiating the score with respect to input gene expression attributes local changes
in typicality to individual genes, yielding per-gene attribution in the input's native coordinates. Thus, a single trained encoder serves simultaneously as a representation, a
typicality score, and an attribution map.

\paragraph{Contributions.}
\begin{itemize}[noitemsep,topsep=2pt]
\item \textbf{A density-estimation strategy for cell states.} We propose a strategy for estimating the relative density of cell states that relies on JEPA-SCORE \citep{gaussian-embeddings}, the implicit density estimator of Gaussian-embedding JEPAs. Combined with biologically meaningful data augmentations and batch-effect controls, JEPA-SCORE yields a relative estimate of cell-state density on the learned manifold, which we read within a cell type as typicality, despite the stochastic, sparse, and batch-confounded nature of the data.
\item \textbf{Differentiable gene-level attribution of density.}
A key novelty of LeCellModel is that the density score is itself differentiable with respect to gene expression. We therefore differentiate it to identify, for each cell, the genes whose local expression changes most affect its JEPA-SCORE, turning a cell-level relative density estimate into direct gene-level attribution. Unlike differential expression, it requires no predefined contrast group.
\item \textbf{LeCellModel: a Gaussian-latent JEPA for single-cell RNA-seq.} We instantiate this idea in a teacher-free, single-network JEPA combining a lightweight MLP backbone, gene-space augmentations, SIGReg~\citep{lejepa-provable}, and adversarial batch correction. The architecture is light enough to train per-dataset and to retain a tractable Jacobian, which is what makes the density readout practical (Section~\ref{sec:train}).
\item \textbf{Cell identity preserved at atlas scale.} LeCellModel preserves cell identity while integrating across batches at atlas scale, leading scIB Total on HLCA Core (584K cells) and Heart v2 (704K cells, 32 preparations), and tying scVI on Pancreas; it also leads GeneJEPA trained on the same data and the foundation models scGPT and UCE (Section~\ref{sec:integration}; App.~\ref{app:genejepa}).
\item \textbf{Density-driven discovery in disease tissue.} JEPA-SCORE recovers stimulated cells in a controlled cytokine panel and tracks pathway-response magnitude. Applied to idiopathic pulmonary fibrosis lung, it surfaces a profibrotic monocyte-derived macrophage program and a CD177\textsuperscript{+} activated-Treg subpopulation previously described only in tumors --- both identified purely from the geometry of the embedding (Sections~\ref{sec:perturbation}--\ref{sec:rare}).
\end{itemize}

\section{Methods}
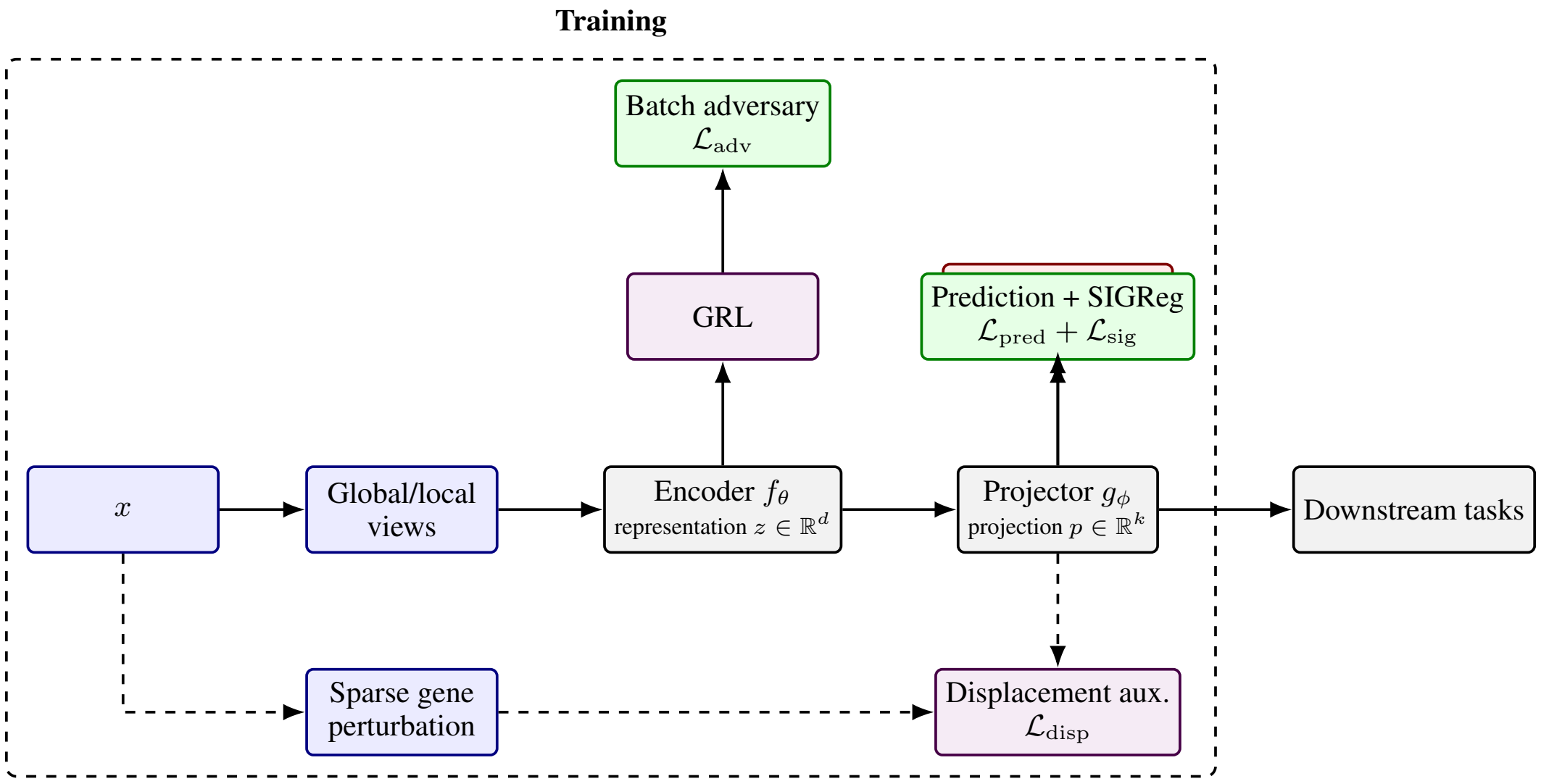
\begin{figure}[t]
\centering
\resizebox{\linewidth}{!}{%
\begin{tikzpicture}[
    >=Latex,
    font=\small,
    node distance=8mm and 10mm,
    arrow/.style={-Latex, thick},
    dashedarrow/.style={-Latex, thick, dashed},
    block/.style={
        draw, rounded corners=2pt, thick,
        minimum height=9mm, minimum width=20mm,
        align=center, inner sep=3pt
    },
    main/.style={block, fill=gray!10},
    input/.style={block, fill=blue!8, draw=blue!50!black},
    loss/.style={block, fill=green!10, draw=green!50!black},
    aux/.style={block, fill=violet!8, draw=violet!55!black},
    scorestyle/.style={block, fill=red!8, draw=red!50!black},
    panel/.style={draw, rounded corners=3pt, dashed, thick, inner sep=6pt},
    stage/.style={font=\bfseries\small}
]

\node[input] (x) {$x$};
\node[input, right=9mm of x] (views) {Global/local\\views};
\node[main, right=11mm of views] (enc) {Encoder $f_\theta$\\{\scriptsize representation $z \in \mathbb{R}^d$}};
\node[main, right=12mm of enc] (proj) {Projector $g_\phi$\\{\scriptsize projection $p \in \mathbb{R}^k$}};
\node[scorestyle, above=12mm of proj] (scorebox) {JEPA-score $S(x)$\\{\scriptsize from $J_F(x)$}};

\node[aux, above=11mm of enc] (grl) {GRL};
\node[loss, above=11mm of grl] (adv) {Batch adversary\\$\mathcal{L}_{\mathrm{adv}}$};
\node[loss, above=11mm of proj] (jepa) {Prediction + SIGReg\\$\mathcal{L}_{\mathrm{pred}}+\mathcal{L}_{\mathrm{sig}}$};

\node[input, below=12mm of views] (pert) {Sparse gene\\perturbation};
\node[aux, below=12mm of proj] (disp) {Displacement aux.\\$\mathcal{L}_{\mathrm{disp}}$};

\node[main, right=14mm of proj] (downstream) {Downstream tasks};

\draw[arrow] (x) -- (views);
\draw[arrow] (views) -- (enc);
\draw[arrow] (enc) -- (proj);
\draw[arrow] (proj) -- (downstream);
\draw[arrow] (proj) -- (scorebox);

\draw[arrow] (enc) -- (grl);
\draw[arrow] (grl) -- (adv);
\draw[arrow] (proj) -- (jepa);

\draw[dashedarrow] (x.south) |- (pert.west);
\draw[dashedarrow] (pert.east) -- (disp.west);
\draw[dashedarrow] (proj) -- (disp);

\node[panel, fit=(x)(views)(enc)(proj)(grl)(adv)(jepa)(pert)(disp)] (trainpanel) {};
\node[stage, above=1mm of trainpanel.north] {Training};

\end{tikzpicture}%
}
\caption{%
\textbf{Overview of LeCellModel.}
Input cells are augmented into global and local views.
The encoder $f_\theta$ maps each view to a representation $z \in \mathbb{R}^d$, and the projector $g_\phi$ maps $z$ to a projection $p \in \mathbb{R}^k$.
During training, batch supervision acts on $z$ through an explicit gradient-reversal layer (GRL), JEPA prediction and SIGReg regularization act on $p$, and an optional displacement auxiliary predicts projection shifts induced by sparse gene-space perturbations applied to the input cell.
The projection $p$ can be used directly as an embedding for downstream tasks, while the JEPA-score is computed separately from the Jacobian of $F=g_\phi\circ f_\theta$.%
}
\label{fig:architecture}
\end{figure}

LeCellModel is built on the Joint Embedding Predictive Architecture (JEPA) framework~\citep{lejepa-provable}.
JEPAs learn representations by requiring augmented views of the same input to be mutually predictable in embedding space --- without reconstructing the input.
For single-cell data, this is natural: since a cell's identity is encoded across many correlated genes, randomly masked and Poisson-resampled views retain enough signal to agree on the same cell state.
In LeCellModel, all augmented views are trained to converge to a shared projection center; no momentum teacher or asymmetric predictor head is used.
SIGReg regularizes the projection distribution toward an isotropic Gaussian --- preventing collapse, enforcing diversity both across embedding dimensions and between cells, and enabling JEPA-score as a density proxy (Section~\ref{sec:score}).
To handle an atlas-scale batch structure, adversarial batch correction penalizes the encoder for retaining batch-predictive features, freeing it from encoding technical variation.
Together, these components yield both reliable cell-state embeddings and a principled density readout (Figure~\ref{fig:architecture}).

\subsection{Problem setup and notation}
\label{sec:setup}

Let $x \in \mathbb{R}^G$ denote a cell's raw expression vector over $G$ highly variable genes, $b \in \{1, \dots, B\}$ its batch label, and $c \in \{1, \dots, C\}$ an optional biological grouping (cell type, or a coarser grouping such as lineage).
We train an encoder $f_\theta : \mathbb{R}^G \to \mathbb{R}^d$ and a projector $g_\phi : \mathbb{R}^d \to \mathbb{R}^k$, and write $F = g_\phi \circ f_\theta$ for their composition, so that $z = f_\theta(x)$ and $p = F(x)$.
An adversary head $h_\psi : \mathbb{R}^d \to \mathbb{R}^B$ predicts batch from the representation.
Encoder, projector, and adversary head are all lightweight two-layer MLPs; full architectures are in App.~\ref{app:arch}.
\label{sec:arch}

\subsection{View generation and augmentation}
\label{sec:views}

Each cell yields $n_g = 1$ global and $n_\ell = 6$ local views through biologically motivated augmentations on raw counts: Poisson resampling (sequencing-depth stochasticity), library-size normalization followed by $\log(1+x)$, and random gene masking at rates $m_g = 0$ (globals) and $m_\ell = 0.6$ (locals, more aggressively perturbed). Poisson resampling is omitted on the atlas-scale benchmarks (HLCA Core, Heart v2). 
The global view is left unmasked so that it serves as an information-rich target for the masked local views (Eq.~\ref{eq:pred}); two global views, masked less than the local views, 
  give similar scIB scores (App.~\ref{app:ablations}).

\subsection{Training objective}
\label{sec:train}

\subsubsection{JEPA prediction loss}

Let $p^g$ denote the projection of the global view and $\{p^\ell_j\}_{j=1}^{n_\ell}$ the projections of the local views. We define the prediction target as $\hat{c} = p^g$ (more generally, $\hat{c} = \tfrac{1}{n_g}\sum_{i=1}^{n_g} p^g_i$ when $n_g > 1$) and minimize
\begin{equation}
\mathcal{L}_{\mathrm{pred}} = \frac{1}{n_g + n_\ell} \sum_{v} \lVert p_v - \hat{c} \rVert^2,
\label{eq:pred}
\end{equation}
averaged over the batch, where the sum runs over all $n_g + n_\ell$ view projections.
Unlike I-JEPA and prior single-cell JEPAs, LeCellModel has no momentum/EMA teacher and no asymmetric predictor head: the same encoder--projector produces both $\hat{c}$ and the predicting views, and gradients flow symmetrically through both with no stop-gradient on $\hat{c}$. Collapse prevention is provided entirely by SIGReg (Eq.~\ref{eq:sig}): without it, a constant projection minimizes $\mathcal{L}_{\mathrm{pred}}$.

\subsubsection{SIGReg regularization}

Following LeJEPA, we regularize the projection distribution toward an isotropic Gaussian $\mathcal{N}(0, I_k)$ using an Epps--Pulley characteristic-function test along $M = 1024$ random one-dimensional projections, evaluated per view-slot and averaged:
\begin{equation}
\mathcal{L}_{\mathrm{sig}} = \frac{1}{(n_g + n_\ell)\,M} \sum_{v} \sum_{m=1}^{M} T_{\mathrm{EP}}^{(v)}(\omega_m).
\label{eq:sig}
\end{equation}
The empirical CF, weighting, truncation, and numerical integration are given in App.~\ref{app:sigreg}. The isotropic Gaussian target also fixes the geometry under which the Jacobian log-volume is read as relative density~\citep{gaussian-embeddings}; replacing SIGReg with VICReg~\citep{vicreg}, which constrains only the first two moments, trains less stably and weakens both scIB and the score (Section~\ref{sec:analysis}).

\subsubsection{Adversarial batch correction}

We train $h_\psi$ to predict the batch label $b$ from the representation through a gradient-reversal layer $\mathrm{GRL}_\lambda$~\citep{grl}:
\begin{equation}
\mathcal{L}_{\mathrm{adv}} = \mathrm{CE}\bigl(h_\psi(\mathrm{GRL}_\lambda(z)), b\bigr),
\label{eq:adv}
\end{equation}
with $\lambda$ ramping from $0$ over a warmup. A conditional variant that preserves coarse biological structure and two stabilizers (spectral normalization on $h_\psi$, label smoothing in cross-entropy) are described in App.~\ref{app:adv}. SIGReg does not integrate batches; removing the adversary lowers Total scIB by $0.09$--$0.20$, with the largest drop in Batch (App.~\ref{app:ablations}).

\subsubsection{Displacement auxiliary}

View prediction primarily constrains the projector along directions induced by the augmentations, leaving sensitivity to other local gene-space perturbations under-determined. We therefore optionally add a displacement auxiliary: a small predictor estimates the projection shift $\Delta p = F(x') - p$ induced by a sparse additive gene-space perturbation, trained with a combined cosine-direction and magnitude loss on a $25\%$ batch subsample after warmup. This provides a modest boost of scIB ($+0.02$ Total on atlas-scale datasets), but regularizing the Jacobian may reduce the precision of the JEPA-score as a typicality signal; the loss form, perturbation procedure, and ablation are in App.~\ref{app:disp}.

\subsubsection{Combined objective}

The final training loss is
\begin{equation}
\mathcal{L} = \mathcal{L}_{\mathrm{pred}} + \lambda_{\mathrm{sig}}\, \mathcal{L}_{\mathrm{sig}} + \lambda_{\mathrm{adv}}\, \mathcal{L}_{\mathrm{adv}} + \lambda_{\mathrm{disp}}\, \mathcal{L}_{\mathrm{disp}}.
\label{eq:total}
\end{equation}
The encoder parameters minimize $\mathcal{L}$ via the GRL; adversary parameters minimize $\mathcal{L}_{\mathrm{adv}}$ directly. The coefficients $\lambda_{\mathrm{adv}}$ and $\lambda_{\mathrm{disp}}$ ramp linearly from $0$ over a warmup; $\lambda_{\mathrm{sig}}$ is fixed. Defaults are in the appendix.

\subsection{JEPA-score}
\label{sec:score}

\textbf{Definition.} For a trained model $F = g_\phi \circ f_\theta$, we define the per-cell score
\begin{equation}
S(x) \;=\; \sum_{i=1}^{k} \log \max\bigl(\sigma_i(J(x)),\, \varepsilon\bigr), \qquad J(x) = \frac{\partial F(x)}{\partial x} \in \mathbb{R}^{k \times G},
\label{eq:score}
\end{equation}
where $\sigma_i(J(x))$ are the singular values of $J(x)$ and $\varepsilon = 10^{-7}$ floors near-zero singular values for numerical stability of the logarithm.
The Jacobian is taken on the composed map $F = g_\phi \circ f_\theta$, i.e.\ with respect to the projector output.

\textbf{Interpretation.} \citet{gaussian-embeddings} showed that, under isotropic Gaussian regularization, the Jacobian log-volume of a trained JEPA approximates $\log p(x)$ at optimality, validated globally on synthetic and image data. Single-cell data departs from the assumptions of this result (sparse counts, masking and resampling views, adversarial batch correction), so we read $S(x)$ as a relative rather than a calibrated log-density. Cells also form a structured manifold of related types with very uneven abundances: across types the score tracks each type's abundance in the dataset, which reflects sampling as well as biology (Section~\ref{sec:analysis}), and the biologically interesting signal is rarely a globally rare cell but a coherent rare \emph{program} inside a known type. We therefore read the score within a cell type, as typicality.

Two mechanisms drive low scores onto rare-and-predictive states. SIGReg's empirical statistic is averaged across the batch, so a sub-population influences it only in proportion to its mixture weight; the rare component is therefore less strongly constrained by the global Gaussian-matching objective. If, additionally, a rare state's cross-view predictability is carried by a compact gene program, the prediction loss is satisfied with few features and points the same way. Rare-and-predictive states therefore concentrate onto few dominant Jacobian directions, yielding a low log-volume score; we confirm this empirically in Section~\ref{sec:analysis}, and Section~\ref{sec:rare} uses the score to localize disease-enriched programs and reads them gene-by-gene through $\nabla_x S$.

\textbf{Computation.} The stabilized score is computed in a single vectorized pass using composable functional transforms (\texttt{vmap}, \texttt{jacrev}, \texttt{svdvals}); details in App.~\ref{app:score-comp}. For a transformer the size of GeneJEPA (over 100M parameters), the per-cell Jacobian costs roughly $2{,}000\times$ more than for our MLP, which makes atlas-scale scoring impractical.

\subsection{Gene-level gradient attribution}
\label{sec:attr}

For a low-scoring cell $x$, we compute the gradient of the score with respect to gene expression,
\begin{equation}
\nabla_x S(x)
\;=\;
\nabla_x \left[
\sum_{i=1}^{k}
\log \max\!\left(
\sigma_i\!\left(\frac{\partial F(x)}{\partial x}\right),
\varepsilon
\right)
\right]
\;\in\;
\mathbb{R}^G,
\label{eq:attr}
\end{equation}
implemented by second-order automatic differentiation, composing the Jacobian (via \texttt{jacrev}) with an outer \texttt{grad} over the stabilized log-volume.

Each entry of $\nabla_x S(x)$ measures the local sensitivity of $S(x)$ to the corresponding gene's expression; negative entries identify genes whose expression locally lowers the score.
\section{Experiments}
\subsection{Datasets and evaluation}

\paragraph{Datasets.}
We evaluate LeCellModel on three integration benchmarks that span orders of magnitude in scale and differ in the dominant source of technical variation:
the Human Lung Cell Atlas (HLCA) Core~\citep{hlca} ($584{,}944$ cells, 14 studies);
the Heart Cell Atlas v2~\citep{heart-ca} ($704{,}296$ cells, 22 donors across 3 sequencing modalities, 32 preparations total);
and a multi-study human pancreas benchmark~\citep{scib} ($16{,}382$ cells, 9 batches spanning six sequencing protocols).
For rare-program discovery we additionally assemble two disease-targeted subsets drawn from the extended HLCA release~\citep{hlca}, combining healthy and IPF donors at roughly $8\%$ disease cells each:
the \emph{HLCA monocyte--macrophage} subset ($257{,}893$ cells, 125 donors, 13 studies, $20{,}313$ IPF cells)
and the \emph{HLCA lymphoid} subset ($159{,}652$ cells, 171 donors, 19 studies, $10{,}913$ IPF cells); subsetting criteria are detailed in Appendix~\ref{app:preprocessing}.
The Parse Biosciences cytokine panel is described in Section~\ref{sec:perturbation}.

\paragraph{Baselines.}
We compare LeCellModel to four standard baselines:
scVI~\citep{scVI}, a variational autoencoder with a negative-binomial likelihood and an explicit batch-conditional decoder;
Harmony~\citep{harmony}, an iterative batch-correction method operating in PCA space;
Scanorama~\citep{scanorama}, a mutual-nearest-neighbor alignment in reduced space;
and raw PCA as an unintegrated lower bound. We also evaluated GeneJEPA~\citep{geneJEPA} (trained from scratch and fine-tuned), scGPT~\citep{scGPT} (zero-shot and fine-tuned) and UCE~\citep{uce} (zero-shot). None was competitive: GeneJEPA scored below PCA, and scGPT and UCE below Harmony; single-run results are in Appendix~\ref{app:genejepa}.

\paragraph{Batch-integration metrics.}
We follow the \texttt{scib\_metrics}~\citep{scib-metrics} reimplementation of scIB~\citep{scib} and report ten metrics:
\emph{bio-conservation} --- Isolated Labels, Leiden~\citep{leiden} NMI, Leiden ARI, Silhouette Label, cLISI;
\emph{batch correction} --- BRAS~\citep{bras}, iLISI, kBET, Graph Connectivity, PCR comparison.
Bio and Batch are the arithmetic means of each group; Total is their scIB-convention weighted average ($0.6$ Bio, $0.4$ Batch). All methods are evaluated across four random seeds; hyperparameters are listed in the appendix.

\subsection{Dataset integrations}
\label{sec:integration}

Table~\ref{tab:scib} reports the Bio-Conservation, Batch-Correction, and Total aggregates for LeCellModel and all baselines on the three benchmark datasets; per-metric breakdowns are in Appendix~\ref{app:scib-full}.

\begin{table}[t]
\centering
\small
\caption{Batch-integration performance across the three benchmark datasets, mean $\pm$ standard deviation across $4$ random seeds (42, 1, 2, 3). scIB aggregates per method's output embedding (definitions in Section~\ref{sec:integration} preamble); bold marks the best per-dataset entry; per-metric tables in Appendix~\ref{app:scib-full}.}
\label{tab:scib}
\setlength{\tabcolsep}{3pt}
\small
\resizebox{\textwidth}{!}{%
\begin{tabular}{l ccc ccc ccc}
\toprule
 & \multicolumn{3}{c}{\textbf{HLCA Core}} & \multicolumn{3}{c}{\textbf{Heart v2}} & \multicolumn{3}{c}{\textbf{Pancreas}} \\
\cmidrule(lr){2-4}\cmidrule(lr){5-7}\cmidrule(lr){8-10}
Method & Bio & Batch & Total & Bio & Batch & Total & Bio & Batch & Total \\
\midrule
PCA        & $0.682${\scriptsize $\pm.008$} & $0.332${\scriptsize $\pm.003$} & $0.542${\scriptsize $\pm.006$} & $0.763${\scriptsize $\pm.006$} & $0.314${\scriptsize $\pm.002$} & $0.583${\scriptsize $\pm.003$} & $0.638${\scriptsize $\pm.000$} & $0.254${\scriptsize $\pm.000$} & $0.485${\scriptsize $\pm.000$} \\
Scanorama  & $0.707${\scriptsize $\pm.000$} & $0.388${\scriptsize $\pm.017$} & $0.580${\scriptsize $\pm.006$} & $0.762${\scriptsize $\pm.000$} & $0.342${\scriptsize $\pm.009$} & $0.594${\scriptsize $\pm.004$} & $0.770${\scriptsize $\pm.004$} & $0.556${\scriptsize $\pm.006$} & $0.684${\scriptsize $\pm.004$} \\
Harmony    & $0.685${\scriptsize $\pm.007$} & $0.516${\scriptsize $\pm.013$} & $0.618${\scriptsize $\pm.009$} & $0.843${\scriptsize $\pm.005$} & $0.554${\scriptsize $\pm.007$} & $0.727${\scriptsize $\pm.004$} & $0.727${\scriptsize $\pm.001$} & $0.587${\scriptsize $\pm.000$} & $0.671${\scriptsize $\pm.001$} \\
scVI       & $\mathbf{0.729}${\scriptsize $\pm.001$} & $0.551${\scriptsize $\pm.006$} & $0.658${\scriptsize $\pm.002$} & $0.818${\scriptsize $\pm.001$} & $0.543${\scriptsize $\pm.005$} & $0.708${\scriptsize $\pm.002$} & $\mathbf{0.831}${\scriptsize $\pm.003$} & $0.594${\scriptsize $\pm.007$} & $0.736${\scriptsize $\pm.003$} \\
\midrule
\textbf{LeCellModel} & $0.727${\scriptsize $\pm.005$} & $\mathbf{0.632}${\scriptsize $\pm.003$} & $\mathbf{0.689}${\scriptsize $\pm.003$} & $\mathbf{0.849}${\scriptsize $\pm.002$} & $\mathbf{0.627}${\scriptsize $\pm.001$} & $\mathbf{0.760}${\scriptsize $\pm.002$} & $0.807${\scriptsize $\pm.010$} & $\mathbf{0.640}${\scriptsize $\pm.007$} & $\mathbf{0.741}${\scriptsize $\pm.008$} \\
\bottomrule
\end{tabular}%
}
\end{table}

\begin{figure}[t]
\centering
\includegraphics[width=\linewidth]{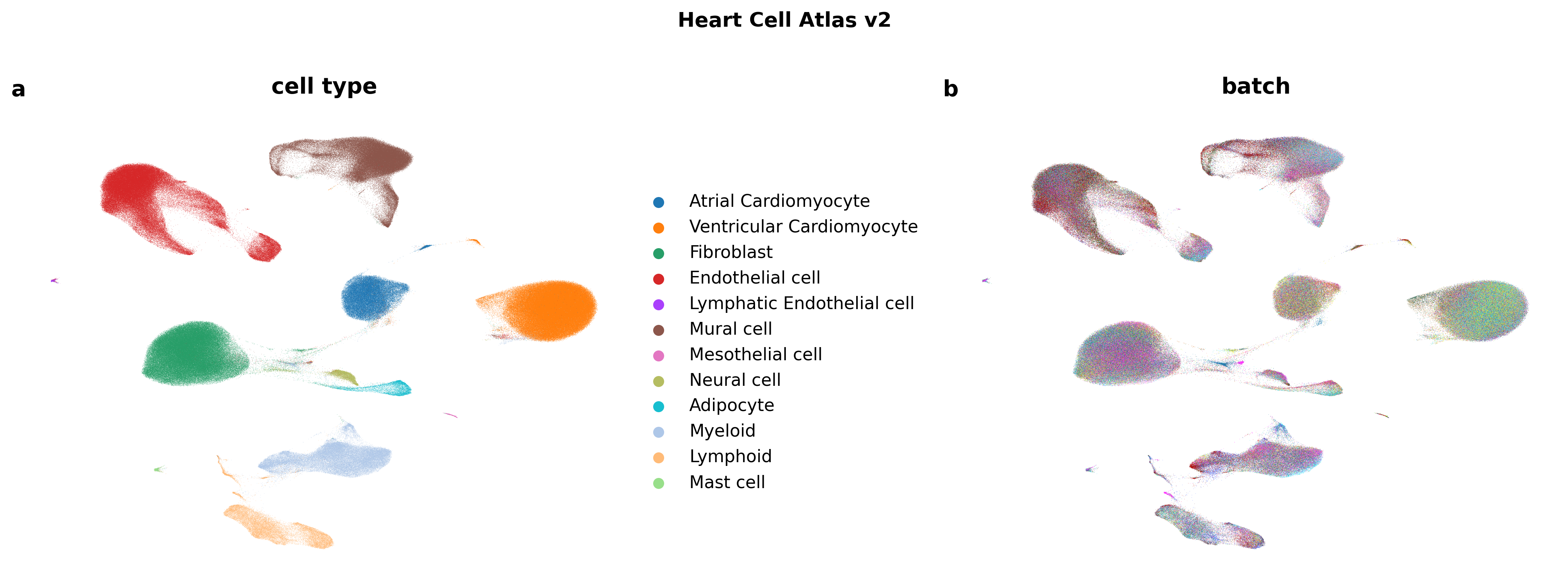}
\caption{\textbf{LeCellModel embedding of the Heart Cell Atlas v2.} UMAP of $704{,}296$ cells encoded by LeCellModel.
\textbf{(a)} Colored by cell type. Twelve major cardiac populations form well-separated clusters, preserving biological structure.
\textbf{(b)} Colored by batch (32 preparations across scRNA-seq, snRNA-seq, and 10x Multiome). Cells from different batches mix within each cluster, demonstrating effective correction of multi-modality batch effects.}
\label{fig:integration-umaps}
\end{figure}

\paragraph{Summary.}
\textbf{LeCellModel attains the highest Total score on all three benchmarks}, with the lead at atlas scale being many times larger than the cross-seed standard deviation: HLCA Core $\Delta = +0.031$ over scVI, Heart v2 $\Delta = +0.033$ over Harmony. On small-scale Pancreas, LeCellModel remains competitive, tying scVI on Total ($0.741 \pm 0.008$ vs.\ $0.736 \pm 0.003$) with LeCellModel leading Batch-Correction and scVI leading Bio-Conservation --- a regime where most atlas-scale methods give up ground. scGPT, UCE and GeneJEPA trail LeCellModel on both Bio and Batch (App.~\ref{app:genejepa}).

\FloatBarrier
\subsection{Validating rare-state detection with ground truth}
\label{sec:perturbation}

The JEPA-score targets unsupervised discovery, so its validation requires a setting where a rarity label exists but is withheld from training. The Parse Biosciences PBMC cytokine-stimulation panel~\citep{parse-pbmc} provides this: stimulated and resting cells are independently labeled by exposure condition --- though within the stimulated population the magnitude and character of the response varies, and not every labeled-stimulated cell exhibits a detectable cytokine signature. We subsample stimulated cells to fractions $f \in \{0.01, 0.03, 0.05, 0.10\}$ of the training mix and train without using the label. We report AUROC and AUPRC against the held-out label, and Spearman~$\rho$ between the score and the canonical pathway signature (STAT3 targets for IL-10; STAT6 for IL-4) within stimulated cells only --- asking whether the score also ranks cells by \emph{degree} of response, not merely stim vs.\ resting.

We evaluate four (cytokine, cell type) pairs spanning a range of effect sizes: IL-10~$\times$~CD16 Mono, IL-10~$\times$~CD14 Mono, IL-4~$\times$~CD14 Mono, and IL-4~$\times$~CD16 Mono (weaker IL-4 response; pair descriptions and cell counts are in Appendix~\ref{app:contam-pairs}). Baselines are seven density- and reconstruction-based anomaly scores --- PCA-kNN, PCA-Mahal, PCA-IForest, scVI-IForest, scVI-kNN, scVI-Mahal, and scVI-recon --- all fit without labels.

\begin{figure}[ht]
\centering
\includegraphics[width=\linewidth]{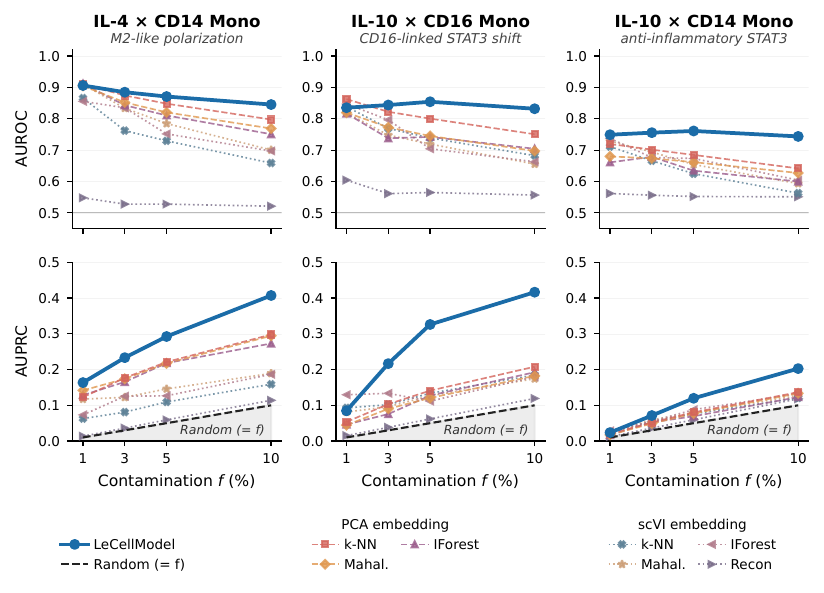}
\caption{\textbf{Rare-state detection benchmark.} Stimulated cells at minority fractions $f \in \{1\%,3\%,5\%,10\%\}$; no label seen during training. AUROC (top) and AUPRC (bottom). Three pairs have established biological responses; the weaker IL-4 response in CD16 monocytes is reported in Appendix~\ref{app:contam-full}.}
\label{fig:contam-panel}
\end{figure}

Each density-based baseline rests on a single geometric assumption, and as $f$ grows, the stimulated minority inflates these density estimates and performance deteriorates substantially (e.g.\ scVI-kNN drops $0.839 \to 0.683$ AUROC on IL-10~$\times$~CD16). The JEPA score remains stable across $f$, leading on all three established-response pairs in both AUROC and AUPRC beyond $f = 1\%$ (Figure~\ref{fig:contam-panel}; Tables~\ref{tab:contam-full},~\ref{tab:contam-auprc}).
Computing the same kinds of scores in LeCellModel's own embedding does not reproduce this: on the three established-response pairs, $k$NN density and Mahalanobis distance fall below
chance, and isolation forest trails JEPA-SCORE, coming close only on IL-4~$\times$~CD14 at low $f$ (Table~\ref{tab:contam-sameemb}).

Within stimulated cells, the JEPA-score correlates with response \emph{magnitude} (Spearman~$\rho$ against the pathway signature, z-scored on resting cells, without access to the gene list); LeCellModel leads on the biologically stronger pairs (Table~\ref{tab:contam-spearman}; pair biology in App.~\ref{app:contam-pairs}).

\FloatBarrier
\subsection{Rare program discovery}
\label{sec:rare}

Within a cell type, low-JEPA-score cells flag rare-and-predictive states (Section~\ref{sec:score}). We can recover the gene program of such a region from two complementary directions: \textbf{gene $\to$ cells}, by finding genes whose expression spatially co-locates with low scores on the embedding $k$NN graph (bivariate Moran's I; App.~\ref{app:gradclust}, Table~\ref{tab:morans-program}), or \textbf{cells $\to$ genes}, by taking a low-score neighborhood and ranking genes by their local effect on the score ($\nabla_x S$; Section~\ref{sec:attr}). The two views converge on the program where they overlap, but $\nabla_x S$ operates per-cell, surfacing score \emph{drivers} rather than co-occurring markers — a distinction no population-level correlation can make. We apply this to the HLCA monocyte--macrophage cohort (13 studies) and the HLCA lymphoid normal$+$IPF cohort (19 studies); batch correction across these contributors is what makes a shared program recoverable.

\paragraph{Monocyte-derived macrophages.}
Within monocyte-derived macrophages, LeCellModel isolates a low-JEPA-score subpopulation corresponding to a profibrotic macrophage program (SPP1, CHI3L1, CHIT1, ADAMDEC1)~\citep{spp1-profibrotic}, consistent with the fibrosis-associated SPP1\textsuperscript{+} state highlighted in the HLCA disease analysis~\citep{hlca} and further extended here by matrix-remodeling and immunosuppressive effectors (CTSK, SIGLEC15). This population is enriched for IPF cells (31.7\% vs.\ 12.5\% baseline within monocyte-derived macrophages; 2.53-fold enrichment), spanning 53 donors (16 IPF).

\begin{figure}[t]
\centering
\includegraphics[width=\linewidth]{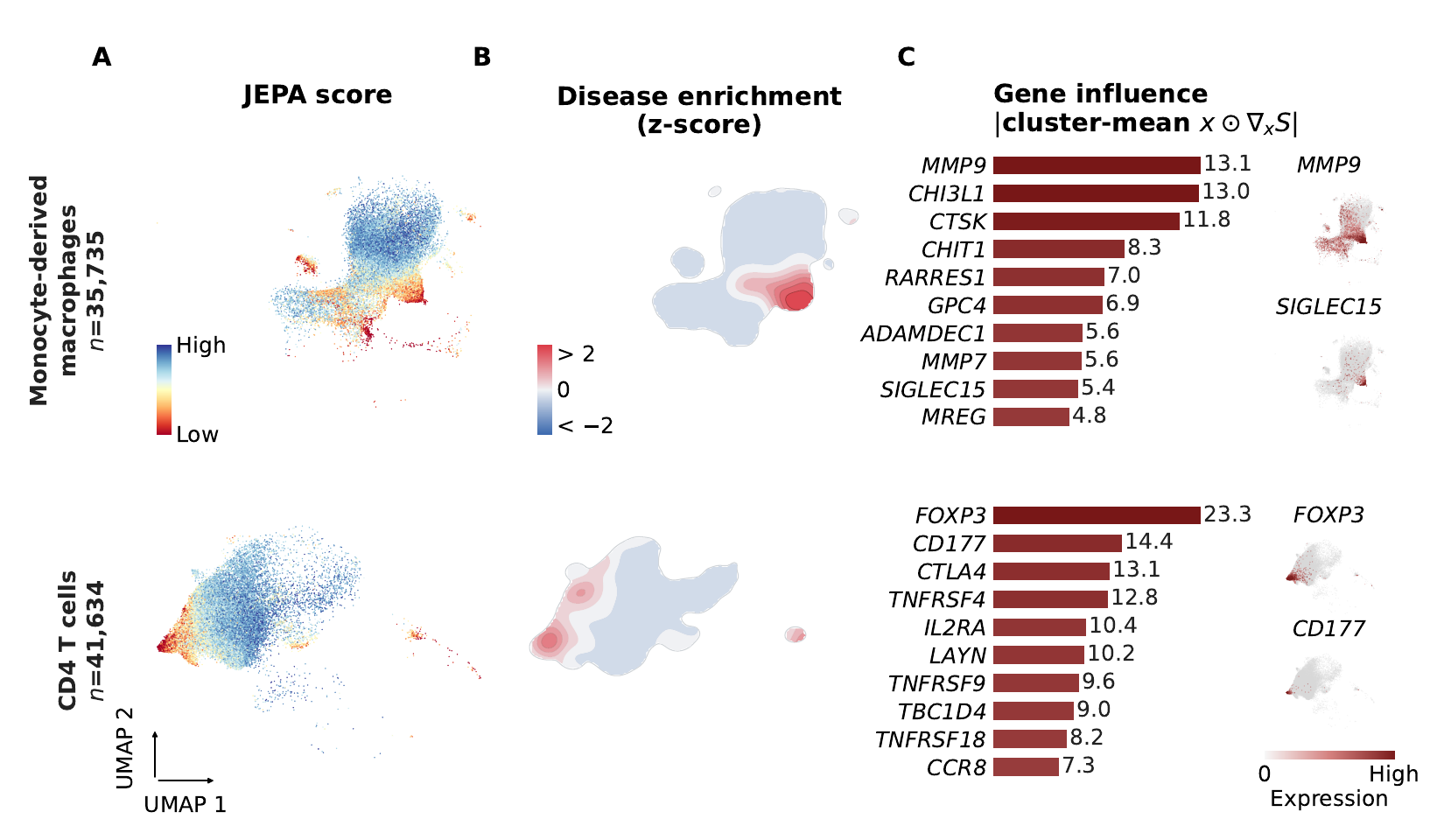}
\caption{
\textbf{Rare-program discovery and score-based gene attribution in IPF lung.}
For each cohort, UMAP is computed on all cells; each row shows the indicated cell type at its coordinates in that full embedding.
\textbf{(A) JEPA score:} Low-JEPA-score cells (red) form a coherent neighborhood within the cell type.
\textbf{(B) disease enrichment:} $k$NN-smoothed disease enrichment $z$-score; the low-JEPA-score region is enriched for IPF cells relative to the cell-type baseline
(monocyte-derived macrophages: 31.7\% vs.\ 12.5\%, $2.53\times$; CD4 T cells: 34.7\% vs.\ 5.3\%, $6.57\times$).
\textbf{(C) gene influence:} Top-10 genes by absolute cluster-mean input$\times$gradient over the low-JEPA-score cells; representative gene-expression UMAPs are shown to the right of
each row.
In monocyte-derived macrophages, the low-JEPA-score region recovers a profibrotic macrophage program, while the gradient prioritizes remodeling and checkpoint genes such as \textit{MMP9}, \textit{CTSK}, and \textit{SIGLEC15}, rather than the most abundant marker, \textit{SPP1}. In CD4 T cells, the same analysis recovers a CD177\textsuperscript{+} activated-Treg program centered on \textit{FOXP3}, \textit{CD177}, \textit{CTLA4}, \textit{CCR8}, and \textit{LAYN}. Cluster selection uses only the JEPA-score and embedding geometry; no gene list is used as input.
}
\label{fig:rare-program}
\end{figure}

\paragraph{CD4 T cells.}
Within CD4 T cells, LeCellModel isolates a low-JEPA-score population defined by FOXP3, CD177, CTLA4, CCR8, and LAYN --- the \emph{CD177\textsuperscript{+} activated-Treg} phenotype previously characterized in tumor microenvironments~\citep{cd177-tumor-tregs}. Its appearance in \emph{non-cancerous}, fibrotic IPF lung is, to our knowledge, previously unreported. LeCellModel identifies it from embedding geometry alone: Tregs appear only in a single research group's per-study labels (Meyer 2019/2021) --- and not at all in HLCA's harmonized vocabulary --- so a label-based analysis of this cohort would likely have missed the population entirely. The recovered program overlaps strongly with the CD177\textsuperscript{+} tumor-infiltrating Treg state described by \citet{cd177-tumor-tregs}, including the highly activated and suppressive phenotype associated with recent tissue infiltration. It is also strongly enriched for IPF cells (34.7\% vs.\ 5.3\% baseline within CD4 T cells; 6.57-fold enrichment), spanning 95 donors (17 IPF). Restricting this cluster to CD177\textsuperscript{+} cells further strengthens the disease association, raising the IPF fraction to 56.9\% (10.76-fold over the CD4 baseline), across 51 donors (13 IPF) (Figure~\ref{fig:rare-program}).

\textbf{Gene-level attribution.}
\label{sec:attr-case}
Given a low-JEPA-score subpopulation $C$, we rank genes by the absolute cluster-mean input$\times$gradient, $\lvert \mathbb{E}_{x\in C}[x \odot \nabla_x S] \rvert$, where the per-cell gradient $\nabla_x S(x) \in \mathbb{R}^G$ is defined in Section~\ref{sec:attr}. Unlike integrated gradients on reconstruction or classification losses, which tend toward cell-type markers, this highlights genes that shape a cell's \emph{local typicality and rare-state geometry} --- extending the JEPA-score from a per-cell anomaly detector to a gene-level interpretation tool without any predefined gene list.
In CD4 T cells, CD177 and CCR8 carry the two largest raw-gradient magnitudes ($\lvert \mathbb{E}_{x\in C}[\partial S/\partial x] \rvert \approx 11.0, 9.4$) --- the score is most locally sensitive to the defining markers of the activated-Treg subpopulation. In monocyte-derived macrophages, the gradient elevates matrix-remodeling and checkpoint effectors (MMP9, CTSK, SIGLEC15), while SPP1 --- the most abundant marker of the recovered program --- falls outside the gradient top: the gradient tends to surface what locally drives the score rather than general markers.
The ranking recovers known pathway targets and complements differential expression: on the CD14 cytokine pairs it places the STAT3 and STAT6 targets F13A1 and ALOX15 first, and in IPF it
ranks the program genes CD177 and SIGLEC15 2nd and 9th, against 25th and 43rd by DE (App.~\ref{app:attr-de}).
The underlying scores are stable across seeds and augmentation settings (mean pairwise per-cell Spearman $\rho = 0.965$ and $0.843$ across four seeds in the macrophage and lymphoid cohorts; App.~\ref{app:stability}).

\section{Analysis}
\label{sec:analysis}

\paragraph{Predictive selectivity.}
At the cell-type scale, the score behaves as the global density reading of \citet{gaussian-embeddings} predicts: on Heart v2, mean JEPA-score increases with cell-type abundance (Spearman~$\rho = 0.86$ between $\log_{10}$ frequency and mean score). Section~\ref{sec:score} predicts a finer mechanism \emph{within} a cell type: under SIGReg, rare states whose cross-view predictability is carried by a compact gene program should concentrate their representation onto fewer dominant directions than common states do. We test this directly: for each population, we measure how well its embedding can be approximated by the one or two best-aligned genes (best-1 and best-2 cosine similarity, $\mathrm{cos}_1$, $\mathrm{cos}_2$).

The prediction holds. For CD177\textsuperscript{+} Tregs (FOXP3\,$>$\,1 and CD177\,$>$\,1, $n = 182$), LeCellModel reaches $\mathrm{cos}_1/\mathrm{cos}_2 = 0.812/0.867$ against $0.367/0.477$ on the broader CD4 T-cell pool ($n=500$); profibrotic monocyte-derived macrophages show the same gap ($0.717/0.821$ vs.\ $0.385/0.534$). The effect is selective to the SIGReg-induced geometry: under scVI, the rare-vs-common contrast almost vanishes (CD177\textsuperscript{+} Treg $0.468/0.593$ vs.\ CD4 $0.359/0.473$; profibrotic monocyte-derived macrophages $0.326/0.465$ vs.\ general monocyte-derived macrophages $0.335/0.450$). The score's geometric mechanism, not just embedding quality, is what makes within-type rare-state recovery work.

\paragraph{Regularizer ablation.} Replacing SIGReg with VICReg~\citep{vicreg}, which constrains only the first two moments, weakens both the score and the representation: the CD177\textsuperscript{+} Treg share in the rarest 1\% falls from $69\%$ to $37\%$, IL-10 AUROC drops by $0.03$ on average, and HLCA Core scIB Total falls from $0.689$ to $0.659$ (App.~\ref{app:ablations}).
\section{Conclusion}
\label{sec:conclusion}

The key design choice in LeCellModel is that SIGReg both prevents collapse and gives the Jacobian log-volume its meaning. The JEPA-score and its gradient therefore come directly from the trained model rather than being added afterward, and they are interpretable because of how the model was trained.

\paragraph{Toward atlas-scale reference embeddings.} A natural next step is to use LeCellModel as an atlas reference: train once on a large multi-study cohort --- potentially with a heavier encoder backbone such as a transformer --- freeze the model, and map query cells into the resulting coordinate system. New samples would inherit an integrated embedding, a per-cell JEPA-score against a fixed reference geometry, and gene-level $\nabla_x S$ attribution, turning rare-state discovery into a query-time operation rather than a per-dataset retraining cycle. An open question is how the score's typicality interpretation transfers when query cells fall outside the training manifold.

\paragraph{Limitations.}
LeCellModel's cross-seed variance on Total scIB is larger in the low-data regime (Pancreas $\sigma = 0.008$) than at the atlas scale (Heart v2, HLCA Core $\sigma \le 0.003$), so per-seed
rankings on small datasets should be read with caution. The displacement-auxiliary ablation (App.~\ref{app:disp}) uses slightly different adversary and SIGReg coefficients than the
with-displacement runs, selected for stability without the auxiliary; the comparison is therefore not perfectly hyperparameter-matched. The JEPA-score is a geometric quantity of a
trained model, not a guaranteed biological readout: it depends on training dynamics and on the augmentations the encoder was exposed to, so runs can disagree at the margin of the rare
tail, though changed cells stay near the cutoff (App.~\ref{app:stability}). We cannot yet separate model uncertainty from genuine rarity or test the gradient against finite count
perturbations, and our negative control (App.~\ref{app:negative-control}) confirms only known phenotypes; new states need replication and experimental validation. Finally, $\nabla_x S$
identifies the genes whose local expression most strongly affects the score; these are interpretive markers of the rare-state geometry, not validated biological drivers.

\begin{ack}
The authors received no specific funding for this work and declare no competing interests.
\end{ack}

\bibliographystyle{plainnat}
\bibliography{references}

\clearpage
\appendix
\section{Data availability}
\label{app:data-availability}

The HLCA Core is available from CZ CELLxGENE (\url{https://cellxgene.cziscience.com/collections/6f6d381a-7701-4781-935c-db10d30de293}); the exact dataset version used here is \texttt{b351804c-293e-4aeb-9c4c-043db67f4540.h5ad}. The HLCA extended atlas is available from the same portal; the monocyte--macrophage and lymphoid subsets used in Section~\ref{sec:rare} are obtained by manually subsetting the extended release to the corresponding lineages and to normal$+$IPF donors only. The Heart Cell Atlas v2 is available at \url{https://www.heartcellatlas.org/}. The multi-study human pancreas benchmark (\texttt{human\_pancreas\_norm\_complexBatch.h5ad}) is distributed by Luecken \emph{et al.} at \url{https://figshare.com/articles/dataset/Benchmarking_atlas-level_data_integration_in_single-cell_genomics_-_integration_task_datasets_Immune_and_pancreas_/12420968}. The Parse Biosciences PBMC cytokine-stimulation panel is available at \url{https://www.parsebiosciences.com/datasets/10-million-human-pbmcs-in-a-single-experiment/}; for the rare-state detection experiments (Section~\ref{sec:perturbation}) we use the $2{,}000$-HVG version distributed by CellFlow at \url{https://cellflow.readthedocs.io/en/stable/user/genapi/cellflow.datasets.pbmc_cytokines.html}.

\section{Dataset preprocessing details}
\label{app:preprocessing}

Raw count matrices are used as input; library-size normalization and \texttt{log1p} are applied inside the view generator (Section~\ref{sec:views}).

\paragraph{HVG selection.}
We use $2{,}000$ HVGs for the three integration benchmarks and $3{,}000$ HVGs for the HLCA monocyte--macrophage and lymphoid subsets, selected with the batch-aware \texttt{seurat\_v3} flavor on raw counts.

\paragraph{HLCA cohort subsetting.}
Both subsets were extracted from the HLCA extended release by filtering to the relevant lineage cell types and normal$+$IPF donors, excluding Kaminski \emph{et al.}~2020 from both cohorts, and dropping datasets contributing fewer than $5{,}000$ (myeloid) or $1{,}000$ (lymphoid) cells.
This yields 13 studies for the monocyte--macrophage cohort and 19 for the lymphoid cohort.

\paragraph{Atlas-scale view generation.}
Poisson resampling is omitted on atlas-scale datasets (HLCA Core, Heart v2) where dataset-level depth variability already provides sufficient view diversity.

\section{Architecture}
\label{app:arch}

The \textbf{encoder} $f_\theta$ is a two-layer MLP with a trailing LayerNorm:
\[
\textsf{Linear}(G \to h) \to \textsf{LayerNorm}(h) \to \textsf{GELU} \to \textsf{Dropout} \to \textsf{Linear}(h \to d) \to \textsf{LayerNorm}(d),
\]
with default $d = 256$, $h = 512$.

The \textbf{projector} $g_\phi$ is a two-layer MLP \emph{without} a trailing norm:
\[
\textsf{Linear}(d \to 512) \to \textsf{LayerNorm}(512) \to \textsf{GELU} \to \textsf{Dropout} \to \textsf{Linear}(512 \to k),
\]
with default $k = 128$.

The \textbf{adversary head} $h_\psi$ is a two-layer MLP with a leading LayerNorm and LeakyReLU activation, producing logits over $B$ batches:
\[
\textsf{LayerNorm}(d) \to \textsf{Linear}(d \to d) \to \textsf{LayerNorm}(d) \to \textsf{LeakyReLU} \to \textsf{Dropout}(0.1) \to \textsf{Linear}(d \to B).
\]
The conditional adversary prepends a learned cell-type embedding $e_c \in \mathbb{R}^{32}$ to the input, replacing the first LayerNorm and Linear with $\textsf{LayerNorm}(d+32) \to \textsf{Linear}(d+32 \to d)$. The conditional variant is not used in any experiment reported in this paper; all results use the unconditional adversary.

\section{View generation details}
\label{app:views}

Global and local view augmentations are applied to raw counts $x \in \mathbb{R}^G$: (i) Poisson resampling, $\tilde{x}_j \sim \mathrm{Poisson}(x_j)$ per gene; (ii) library-size normalization to target sum $10^4$ followed by $\log(1+\cdot)$; (iii) random gene masking at rate $m_g$ (global) or $m_\ell$ (local). $(m_g, m_\ell)$ are $(0, 0.6)$ for all three integration benchmarks; the IPF and rare-state detection experiments use $(0.5, 0.7)$ (see Tables~\ref{tab:lejac-hparams-ipf} and~\ref{tab:lejac-hparams-contam}).

\section{SIGReg details}
\label{app:sigreg}

Let $\{p_n\}_{n=1}^{N} \subset \mathbb{R}^k$ denote the projections of a single view-slot across a batch of size $N$, with $p \in \mathbb{R}^k$ as defined in Section~\ref{sec:setup}.

\paragraph{Random directions.} We draw $M = 1024$ unit directions $\{\omega_m\}_{m=1}^{M} \subset S^{k-1}$ uniformly on the unit sphere (resampled each step) and reduce isotropy in $\mathbb{R}^k$ to $M$ univariate normality tests on the scalar projections $u_n^{(m)} = \omega_m^{\top} p_n$.

\paragraph{Empirical characteristic function.} For each direction $\omega_m$, the empirical CF of the scalar projections $\{u_n^{(m)}\}_{n=1}^{N}$ is
\begin{equation}
\hat{\varphi}_{\omega_m}(t) \;=\; \frac{1}{N}\sum_{n=1}^{N} \exp\!\big(i\, t\, u_n^{(m)}\bigr).
\label{eq:ecf}
\end{equation}

\paragraph{Epps--Pulley statistic.} We compare $\hat{\varphi}_{\omega_m}$ to the standard-normal CF $\varphi_{\mathcal{N}}(t) = \exp(-t^2/2)$ through a weighted squared-error integral,
\begin{equation}
T_{\mathrm{EP}}(\omega_m) \;=\; N \int_{-T}^{T} \bigl| \hat{\varphi}_{\omega_m}(t) - \varphi_{\mathcal{N}}(t) \bigr|^2 \, w(t)\, dt,
\label{eq:tep}
\end{equation}
with weighting $w(t) = \exp(-t^2/2)$ and truncation $T = 5$; the integral is approximated by a 17-point trapezoidal rule on a uniform grid over $[-T, T]$.

\paragraph{Aggregation across view-slots.} At each training step we evaluate Eq.~\ref{eq:tep} once per view-slot ($n_g + n_\ell$ evaluations) and average across view-slots and directions to obtain $\mathcal{L}_{\mathrm{sig}}$ in Eq.~\ref{eq:sig}. Per-view-slot evaluation preserves the i.i.d.\ assumption across $\{p_n\}_{n=1}^{N}$ that pooling projections across view-slots would violate.

\section{Adversarial batch correction details}
\label{app:adv}

\textbf{Conditional variant.} We condition the adversary on a biological grouping $c$ (cell type, or a coarser grouping such as lineage --- chosen per dataset):
\begin{equation}
\mathcal{L}_{\mathrm{adv}}^{\mathrm{cond}} = \mathrm{CE}\bigl(h_\psi([\mathrm{GRL}_\lambda(z)\,;\, e_c]), b\bigr),
\label{eq:adv-cond}
\end{equation}
where $e_c$ is the learned group embedding defined in App.~\ref{app:arch}. This removes batch effects \emph{within} each group while preserving biological structure \emph{between} groups, useful when high-confidence biological grouping is available. The conditional variant is not used in any experiment reported in this paper; all results use the unconditional adversary.

\textbf{Stabilization.} Adversarial training with a growing $\lambda$ schedule is prone to instability. We apply two standard stabilizers: spectral normalization on $h_\psi$'s linear layers (bounding its Lipschitz constant) and label smoothing in the cross-entropy (softening overconfident batch predictions). Both are applied in all experiments; label smoothing uses a value of $0.1$.

\section{JEPA-score computation}
\label{app:score-comp}

We compute the stabilized score (Eq.~\ref{eq:score}) in a single vectorized pass. For a minibatch of $N$ cells, reverse-mode differentiation mapped over cells (\texttt{vmap}, \texttt{jacrev}) gives the stacked Jacobians, and a batched singular-value decomposition (\texttt{svdvals}) gives their singular values:
\[
\mathbf{J} = \bigl(J(x_1), \dots, J(x_N)\bigr) \in \mathbb{R}^{N \times k \times G}, \qquad
S(x_n) = \sum_{i=1}^{k} \log \max\bigl(\sigma_i(\mathbf{J}_n),\, \varepsilon\bigr), \quad n = 1, \dots, N.
\]
The floor is a numerical stabilization for near-degenerate singular directions and does not alter the intended geometric interpretation away from that regime.

\section{Displacement auxiliary details}
\label{app:disp}

Let $F = g \circ f$ denote the composed encoder--projector and $R \in \mathbb{R}^{G \times r}$ a fixed random matrix (seeded; held constant across training) with $\ell_2$-normalized columns, where $r = 64$. The auxiliary is applied to a $25\%$ subsample of each batch after warmup. For each cell $x$ (log-normalized counts), the procedure is:

\begin{algorithm}[H]
\small
\caption{Displacement auxiliary (per cell $x \in \mathbb{R}^G$; per-gene std of log-normalized counts $s$; scale $\sigma$; subset size $n_s$). $\mathrm{sg}(\cdot)$ denotes stop-gradient.}
\begin{algorithmic}[1]
\State Sample gene subset $S \subset \{1,\dots,G\}$ uniformly at random, $|S|=n_s$; form indicator $v \in \{0,1\}^G$
\State Sample $\alpha \sim \mathcal{N}(0, \sigma^2)$ \Comment{scalar, per cell}
\State $x' \gets \max(x + \alpha \cdot (v \odot s),\, 0)$ \Comment{scaled by per-gene batch std}
\State $p \gets F(x)$ \Comment{with gradient}
\State $p' \gets F(x')$ \Comment{no gradient}
\State $\Delta p \gets \mathrm{sg}(p') - \mathrm{sg}(p)$;\quad $\bar v \gets v / \lVert v \rVert$;\quad $u \gets R^\top \bar v \,/\, \lVert R^\top \bar v \rVert$
\State $\widehat{\Delta p} \gets h_\theta(p, u)$ \Comment{predictor input: $p$ with gradient}
\end{algorithmic}
\end{algorithm}

$h_\theta$ takes the concatenation $[p;\, u] \in \mathbb{R}^{k+r}$ and outputs $\widehat{\Delta p} \in \mathbb{R}^k$ via
\[
\textsf{Linear}(k+r \to 256) \to \textsf{LayerNorm}(256) \to \textsf{GELU} \to \textsf{Linear}(256 \to k).
\]
We use $n_s = 25$ and $\sigma = 1.0$.

The loss combines a cosine-direction term and a scalar magnitude term, evaluated only on cells with $\lVert\Delta p\rVert > \tau$ ($\tau = 10^{-4}$):
\begin{equation}
\mathcal{L}_{\mathrm{disp}} = \mathbb{E}_{x,v}\!\left[\, 1 - \cos\bigl(\widehat{\Delta p},\, \Delta p\bigr) \;+\; \gamma \cdot \bigl(\lVert\widehat{\Delta p}\rVert - \lVert\Delta p\rVert\bigr)^2 \,\right], \qquad \gamma = 0.1.
\label{eq:disp}
\end{equation}
The entire target $\Delta p$ is detached, so the predictor matches it but the encoder does not shrink displacements to minimize it. Gradient flows into $F$ only through $p$ in $\widehat{\Delta p} = h_\theta(p, u)$; $p'$ is computed without gradient, so the objective pressures the encoder to make small gene-space perturbations predictable in the latent space.

\paragraph{Ablation: scIB without the displacement auxiliary.}
Table~\ref{tab:disp-ablation} reports LeCellModel with and without the displacement auxiliary on the three integration benchmarks. Removing the auxiliary lowers Total on HLCA Core and Heart v2; on Pancreas, Total is essentially unchanged (a slight Bio gain offsets a Batch decline). In the main text, we report the configuration with the auxiliary enabled. The no-displacement runs use slightly different coefficients, selected for stability without the auxiliary: $\lambda_{\mathrm{sig}} = 0.4$ (vs.\ $0.5$); adversary $\lambda$ progress factor $1.6$ (HLCA Core) / $1.6$ (Heart v2) / $1.3$ (Pancreas) vs.\ $1.8/1.8/1.5$; and Pancreas adversary weight $0.8$ (vs.\ $1.0$).

\begin{table}[!ht]
\centering
\small
\setlength{\tabcolsep}{4pt}
\caption{LeCellModel with and without the displacement auxiliary across the three integration benchmarks. Baseline row from Table~\ref{tab:scib} (mean across $4$ seeds); ablation row is single-seed (seed 42).}
\label{tab:disp-ablation}
\begin{tabular}{l ccc ccc ccc}
\toprule
 & \multicolumn{3}{c}{\textbf{HLCA Core}} & \multicolumn{3}{c}{\textbf{Heart v2}} & \multicolumn{3}{c}{\textbf{Pancreas}} \\
\cmidrule(lr){2-4}\cmidrule(lr){5-7}\cmidrule(lr){8-10}
Configuration & Bio & Batch & Total & Bio & Batch & Total & Bio & Batch & Total \\
\midrule
Baseline (w/ disp.)        & 0.727 & 0.632 & 0.689 & 0.849 & 0.627 & 0.760 & 0.807 & 0.640 & 0.741 \\
\quad No displacement aux. & 0.716 & 0.603 & 0.671 & 0.845 & 0.577 & 0.738 & 0.820 & 0.629 & 0.744 \\
\bottomrule
\end{tabular}
\end{table}

\section{Additional ablations}
\label{app:ablations}

We ablate two design choices on all three integration benchmarks against the main-paper configuration: removing the adversarial batch-correction head, and changing the view-augmentation regime to $n_g = 2$ globals with mask rates $m_g = 0.5, m_\ell = 0.7$ (instead of $n_g = 1$, $m_g = 0$, $m_\ell = 0.6$). The baseline row reproduces the multi-seed Total from Table~\ref{tab:scib}; ablation rows are single-seed (seed 42).

\begin{table}[!ht]
\centering
\small
\setlength{\tabcolsep}{4pt}
\caption{LeCellModel ablations across the three integration benchmarks. Baseline row from Table~\ref{tab:scib} (mean across $4$ seeds); ablation rows are single-seed (seed 42).}
\label{tab:ablation-scib}
\begin{tabular}{l ccc ccc ccc}
\toprule
 & \multicolumn{3}{c}{\textbf{HLCA Core}} & \multicolumn{3}{c}{\textbf{Heart v2}} & \multicolumn{3}{c}{\textbf{Pancreas}} \\
\cmidrule(lr){2-4}\cmidrule(lr){5-7}\cmidrule(lr){8-10}
Configuration & Bio & Batch & Total & Bio & Batch & Total & Bio & Batch & Total \\
\midrule
Baseline                       & 0.727 & 0.632 & 0.689 & 0.849 & 0.627 & 0.760 & 0.807 & 0.640 & 0.741 \\
\quad No adversary             & 0.685 & 0.472 & 0.600 & 0.727 & 0.446 & 0.614 & 0.656 & 0.371 & 0.542 \\
\quad $n_g=2$, masks $0.5/0.7$ & 0.714 & 0.619 & 0.676 & 0.849 & 0.596 & 0.748 & 0.794 & 0.637 & 0.731 \\
\bottomrule
\end{tabular}
\end{table}

The adversary is the dominant single-component contributor across all three datasets: removing it lowers Total by $0.089$, $0.146$, and $0.199$ on HLCA Core, Heart v2, and Pancreas, respectively, through both Batch and Bio, with the larger drop in Batch. The alternative view regime is roughly Total-neutral on Heart v2 ($-0.012$) and gives small regressions on HLCA Core ($-0.013$) and Pancreas ($-0.010$); the lighter $n_g = 1$ regime is retained in the main paper.

\paragraph{Regularizer ablation.} For VICReg we set the adversary progress factor to $1.0$ (vs.\ $1.6$ for SIGReg), as training was unstable at $1.6$; all other settings follow the Figure~\ref{fig:rare-program} configuration (seed 1; VICReg $\lambda = 1.0$). The CD177\textsuperscript{+} Treg program here is defined as CD4 T cells with FOXP3\,$>$\,2 and CD177\,$>$\,2 ($n = 162$); its median score percentile moves from $0.55\%$ under SIGReg (seed 1; $69.1\%$ in the rarest 1\%) to $1.49\%$ under VICReg ($37.0\%$). On the cytokine panel, compared at matched seeds (1--2), VICReg lowers AUROC by $0.03$ on average (up to $0.06$) on both IL-10 pairs and changes it by at most $0.022$ on the IL-4 pairs. On HLCA Core, VICReg reaches scIB Total $0.659$ (mean of seeds 42, 1, 2 at $\lambda = 2.0$). The same Jacobian score computed on scVI's encoder stays near chance (AUROC $0.50$--$0.61$ at $f = 5\%$ on all four pairs, seed 42, against $0.71$--$0.87$ for LeCellModel), so the readout depends on the SIGReg-shaped representation.

\section{Training hyperparameters}
\label{app:hparams}

\paragraph{LeCellModel.} Across all datasets we use a two-layer MLP encoder and projector with $d = 256$, $k = 128$, AdamW ($\mathrm{lr} = 5 \times 10^{-4}$, weight decay $10^{-2}$), OneCycle LR with $10$-epoch warmup, gradient clipping at norm $10$, and bf16 mixed precision. Spectral normalization on the adversary and label smoothing ($0.1$) are enabled throughout. Per-dataset settings are in Table~\ref{tab:lejac-hparams}.

\paragraph{Random seeds.} Integration benchmark results (Table~\ref{tab:scib}) are averaged over seeds $\{42, 1, 2, 3\}$ — the framework default and the three smallest positive integers.

\begin{table}[!ht]
\centering
\small
\setlength{\tabcolsep}{5pt}
\caption{LeCellModel training configuration per dataset. ``Disp.'' is the displacement auxiliary. Epochs$=150$ for all datasets.}
\label{tab:lejac-hparams}
\begin{tabular}{lccc}
\toprule
                                      & HLCA Core & Heart v2 & Pancreas \\
\midrule
Batch size                            & 1024 & 1024 & 512  \\
$\lambda_{\mathrm{sig}}$              & 0.5  & 0.5  & 0.5  \\
Adversary weight                      & 0.5  & 0.5  & 1.0  \\
Adv.\ $\lambda$ progress factor       & 1.8  & 1.8  & 1.5  \\
Adv.\ warmup epochs                   & 0    & 0    & 0    \\
$n_g / n_\ell$                        & 1 / 6 & 1 / 6 & 1 / 6 \\
Mask rates $m_g / m_\ell$             & 0 / 0.6 & 0 / 0.6 & 0 / 0.6 \\
Poisson resampling                    & off & off & on  \\
Projector dim $k$                     & 128 & 128 & 128 \\
Disp.\ (weight)                       & 2.0 & 2.0 & 2.0 \\
\bottomrule
\end{tabular}
\end{table}

\paragraph{Rare-program discovery cohorts.} The HLCA monocyte--macrophage and HLCA lymphoid (normal+IPF) subsets used in Section~\ref{sec:rare} are trained with the configuration in Table~\ref{tab:lejac-hparams-ipf}. They share the same encoder architecture (App.~\ref{app:arch}) but use a larger projection dimension ($k = 256$) and a different view-augmentation regime ($n_g = 2$ globals, mask rates $0.5 / 0.7$) than the integration benchmarks; both also use $3{,}000$ HVGs (App.~\ref{app:preprocessing}).

\begin{table}[!ht]
\centering
\small
\setlength{\tabcolsep}{5pt}
\caption{LeCellModel training configuration for the IPF rare-program discovery cohorts (Section~\ref{sec:rare}). Conventions match Table~\ref{tab:lejac-hparams}.}
\label{tab:lejac-hparams-ipf}
\begin{tabular}{lcc}
\toprule
                                      & HLCA mono--mac & HLCA lymphoid \\
\midrule
Batch size                            & 1024 & 1024 \\
$\lambda_{\mathrm{sig}}$              & 0.5  & 0.5 \\
Adversary weight                      & 0.7  & 0.5 \\
Adv.\ $\lambda$ progress factor       & 1.6  & 1.6 \\
Adv.\ warmup epochs                   & 0    & 0   \\
$n_g / n_\ell$                        & 2 / 6 & 2 / 6 \\
Mask rates $m_g / m_\ell$             & 0.5 / 0.7 & 0.5 / 0.7 \\
Poisson resampling                    & on & on \\
Projector dim $k$                     & 256 & 256 \\
Disp.\ (weight)                       & off & off \\
\bottomrule
\end{tabular}
\end{table}

\paragraph{Rare-state detection experiments (Section~\ref{sec:perturbation}).} The four cytokine~$\times$~cell-type runs use the configuration in Table~\ref{tab:lejac-hparams-contam}. Each run trains LeCellModel from scratch on a contaminated mix at one of $f \in \{0.01, 0.03, 0.05, 0.10\}$ stim fractions, with no displacement auxiliary; the adversary uses donor as the batch label.

\begin{table}[!ht]
\centering
\small
\setlength{\tabcolsep}{5pt}
\caption{LeCellModel training configuration for the cytokine rare-state detection benchmark (Section~\ref{sec:perturbation}). Identical across all minority fractions; adv.\ $\lambda$ progress factor differs between CD14 and CD16 pairs.}
\label{tab:lejac-hparams-contam}
\begin{tabular}{lc}
\toprule
                                      & Rare-state detection \\
\midrule
Batch size                            & 1024 \\
$\lambda_{\mathrm{sig}}$              & 0.35 \\
Adversary weight                      & 0.4 \\
Adv.\ $\lambda$ progress factor       & 1.0 (CD16 pairs); 1.4 (CD14 pairs) \\
Adv.\ warmup epochs                   & 5 \\
$n_g / n_\ell$                        & 2 / 6 \\
Mask rates $m_g / m_\ell$             & 0.5 / 0.7 \\
Poisson resampling                    & on \\
Projector dim $k$                     & 256 \\
Disp.\ (weight)                       & off \\
\bottomrule
\end{tabular}
\end{table}

\paragraph{Baselines.} scVI is trained with \texttt{n\_layers} $= 2$, \texttt{n\_latent} $= 40$, \texttt{n\_hidden} $= 256$ for up to $200$ epochs with early stopping. Harmony and Scanorama are run on the standard scanpy log-normalization pipeline ($10^4$ target sum, log1p; Harmony additionally scales to unit variance at max value $10$ and takes the top-$50$ PCs as input); both use library defaults. PCA uses $50$ components on log-normalized, scaled expression.

\paragraph{Compute.} All LeCellModel training runs use a single GPU per job, drawn from a shared institutional cluster with a mixed pool: NVIDIA A40, NVIDIA A100 (40\,GB and 80\,GB PCIe), and NVIDIA H100 NVL. Training and inference both use bf16 mixed precision. A full $150$-epoch training run takes roughly $1$~hour for the atlas-scale benchmarks (HLCA Core, Heart v2) and roughly $3$~minutes for Pancreas; exact wall-clock varies with GPU type. The reported runs represent the final configurations; hyperparameter development required additional runs of comparable scale.

\section{GeneJEPA and foundation-model baselines}
\label{app:genejepa}

\begin{figure}[!ht]
\centering
\includegraphics[width=0.55\linewidth]{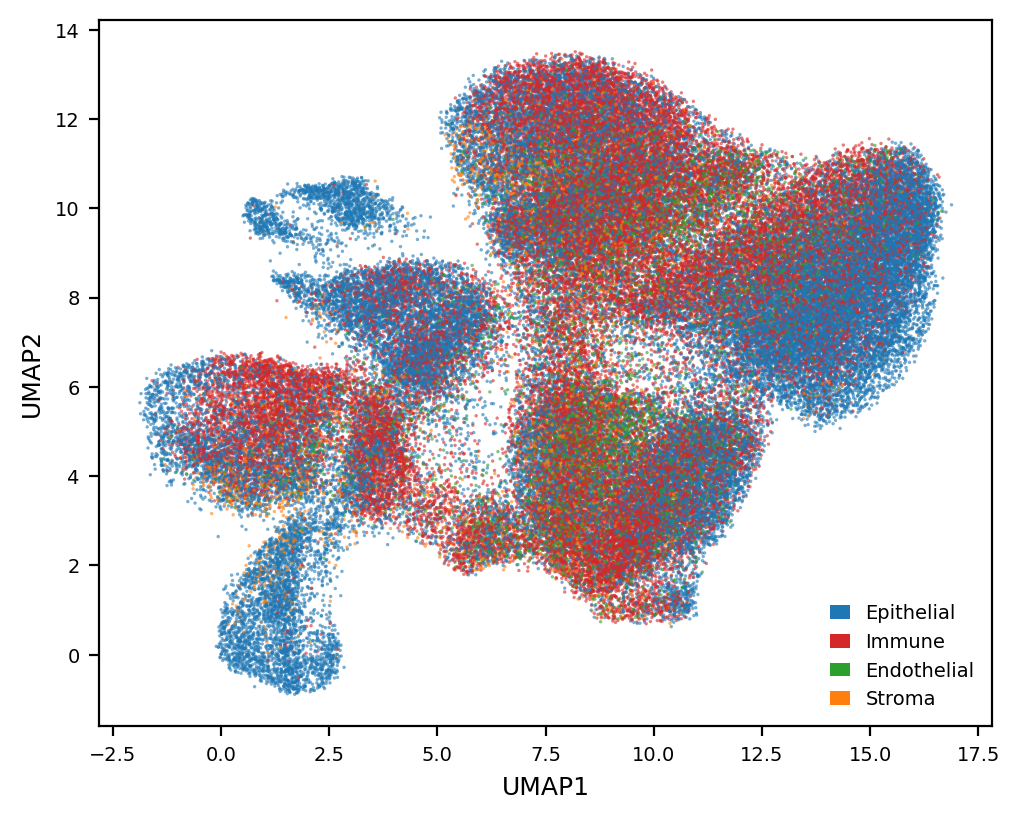}
\caption{\textbf{GeneJEPA zero-shot embedding of HLCA Core, colored by cell type.} The frozen encoder does not produce separated clusters on primary tissue, consistent with the Tahoe-100M $\to$ HLCA domain gap.}
\label{fig:genejepa-umap}
\end{figure}

We evaluate GeneJEPA~\citep{geneJEPA} zero-shot (published Tahoe-100M checkpoint) and, with our own training scripts, fine-tuned and trained from scratch on our benchmarks. All three settings trail LeCellModel and do not separate cell types (Leiden NMI $0.06$--$0.24$); Figure~\ref{fig:genejepa-umap} shows the zero-shot UMAP on HLCA Core. We also report scGPT~\citep{scGPT} (zero-shot and fine-tuned) and UCE~\citep{uce} (zero-shot, 4-layer model). UCE's $1280$-dimensional embedding makes scIB's silhouette metrics infeasible on the full HLCA Core, so it is scored on a stratified $100$k-cell subsample, with LeCellModel scored on the identical cells. Table~\ref{tab:scib-genejepa} gives the per-metric breakdown for zero-shot GeneJEPA and Table~\ref{tab:fm-baselines} the aggregates for all settings.

\begin{table}[!ht]
\centering
\scriptsize
\setlength{\tabcolsep}{3pt}
\caption{GeneJEPA zero-shot scIB on HLCA Core ($584{,}944$ cells), single run. LeCellModel row reproduced from Table~\ref{tab:scib-lung} for reference.}
\label{tab:scib-genejepa}
\resizebox{\textwidth}{!}{%
\begin{tabular}{l ccccc c ccccc c c}
\toprule
 & \multicolumn{6}{c}{\textbf{Bio-Conservation}} & \multicolumn{6}{c}{\textbf{Batch-Correction}} & \\
\cmidrule(lr){2-7}\cmidrule(lr){8-13}
Method & Iso.\ Lab.\ & NMI & ARI & Sil.\ & cLISI & \textbf{Bio} & BRAS & iLISI & kBET & GC & PCR & \textbf{Batch} & \textbf{Total} \\
\midrule
GeneJEPA (zero-shot) & $0.450$ & $0.090$ & $0.031$ & $0.440$ & $0.935$ & $0.389$ & $0.793$ & $0.231$ & $0.205$ & $0.089$ & $0.685$ & $0.401$ & $0.394$ \\
\textbf{LeCellModel} & $0.617$ & $0.804$ & $0.682$ & $0.533$ & $1.000$ & $0.727$ & $0.815$ & $0.119$ & $0.399$ & $0.905$ & $0.921$ & $0.632$ & $0.689$ \\
\bottomrule
\end{tabular}%
}
\end{table}

\begin{table}[!ht]
\centering
\small
\setlength{\tabcolsep}{4pt}
\caption{GeneJEPA and foundation-model baselines, scIB aggregates, single run each. LeCellModel row from Table~\ref{tab:scib} (mean over four seeds). $^{*}$UCE on HLCA Core is scored on a stratified $100$k-cell subsample, against LeCellModel on the same cells.}
\label{tab:fm-baselines}
\begin{tabular}{l ccc ccc}
\toprule
 & \multicolumn{3}{c}{\textbf{HLCA Core}} & \multicolumn{3}{c}{\textbf{Pancreas}} \\
\cmidrule(lr){2-4}\cmidrule(lr){5-7}
Method & Bio & Batch & Total & Bio & Batch & Total \\
\midrule
GeneJEPA (zero-shot)          & 0.389 & 0.401 & 0.394 & 0.389 & 0.381 & 0.386 \\
GeneJEPA (from scratch)       & 0.466 & 0.507 & 0.482 & 0.403 & 0.158 & 0.305 \\
GeneJEPA (fine-tuned)         & 0.386 & 0.396 & 0.390 & 0.386 & 0.377 & 0.382 \\
scGPT (zero-shot)             & 0.598 & 0.487 & 0.554 & 0.682 & 0.490 & 0.605 \\
scGPT (fine-tuned)            & 0.706 & 0.414 & 0.589 & 0.760 & 0.496 & 0.655 \\
UCE (zero-shot)               & 0.698$^{*}$ & 0.349$^{*}$ & 0.558$^{*}$ & 0.600 & 0.373 & 0.510 \\
\midrule
LeCellModel                   & 0.727 & 0.632 & 0.689 & 0.807 & 0.640 & 0.741 \\
LeCellModel (100k subsample)  & 0.732$^{*}$ & 0.664$^{*}$ & 0.705$^{*}$ & -- & -- & -- \\
\bottomrule
\end{tabular}
\end{table}

\FloatBarrier
\section{Full per-metric scIB tables}
\label{app:scib-full}

Tables~\ref{tab:scib-lung}--\ref{tab:scib-pancreas} report the ten individual scIB metrics for each method on each benchmark. Bio-Conservation is the mean of Isolated Labels, Leiden NMI, Leiden ARI, Silhouette Label, and cLISI; Batch-Correction is the mean of BRAS (Batch-label Adjusted Silhouette, a cell-type-stratified silhouette score that measures within-label batch mixing; \texttt{scib\_metrics.bras}), iLISI, kBET, Graph Connectivity, and PCR comparison; Total is the scIB-convention weighted average ($0.6$ Bio $+\,0.4$ Batch). Column aggregates match Table~\ref{tab:scib} up to rounding. 

\begin{table}[!ht]
\centering
\scriptsize
\setlength{\tabcolsep}{3pt}
\caption{HLCA Core ($584{,}944$ cells): per-metric scIB results, mean $\pm$ std across $4$ seeds. Iso.\ Lab.\ = Isolated Labels; NMI/ARI on Leiden; Sil.\ = Silhouette Label; GC = Graph Connectivity; PCR = PCR comparison.}
\label{tab:scib-lung}
\resizebox{\textwidth}{!}{%
\begin{tabular}{l ccccc c ccccc c c}
\toprule
 & \multicolumn{6}{c}{\textbf{Bio-Conservation}} & \multicolumn{6}{c}{\textbf{Batch-Correction}} & \\
\cmidrule(lr){2-7}\cmidrule(lr){8-13}
Method & Iso.\ Lab.\ & NMI & ARI & Sil.\ & cLISI & \textbf{Bio} & BRAS & iLISI & kBET & GC & PCR & \textbf{Batch} & \textbf{Total} \\
\midrule
PCA            & $0.556${\scriptsize $\pm.001$} & $0.763${\scriptsize $\pm.015$} & $0.545${\scriptsize $\pm.023$} & $0.549${\scriptsize $\pm.003$} & $1.000$ & $0.682${\scriptsize $\pm.008$} & $0.634${\scriptsize $\pm.002$} & $0.001${\scriptsize $\pm.001$} & $0.144${\scriptsize $\pm.006$} & $0.881${\scriptsize $\pm.008$} & $0.000$ & $0.332${\scriptsize $\pm.003$} & $0.542${\scriptsize $\pm.006$} \\
Scanorama      & $0.562${\scriptsize $\pm.001$} & $0.795${\scriptsize $\pm.002$} & $0.635$ & $0.545$ & $1.000$ & $0.707$ & $0.778${\scriptsize $\pm.002$} & $0.019$ & $0.199${\scriptsize $\pm.001$} & $0.791${\scriptsize $\pm.003$} & $0.153${\scriptsize $\pm.088$} & $0.388${\scriptsize $\pm.017$} & $0.580${\scriptsize $\pm.006$} \\
Harmony        & $0.538$ & $0.755${\scriptsize $\pm.005$} & $0.598${\scriptsize $\pm.028$} & $0.537$ & $0.999$ & $0.685${\scriptsize $\pm.007$} & $0.812${\scriptsize $\pm.001$} & $0.063$ & $0.318${\scriptsize $\pm.001$} & $0.830${\scriptsize $\pm.001$} & $0.558${\scriptsize $\pm.065$} & $0.516${\scriptsize $\pm.013$} & $0.618${\scriptsize $\pm.009$} \\
scVI           & $0.605${\scriptsize $\pm.004$} & $0.810${\scriptsize $\pm.001$} & $0.707${\scriptsize $\pm.004$} & $0.521${\scriptsize $\pm.001$} & $1.000$ & $\mathbf{0.729}${\scriptsize $\pm.001$} & $0.822${\scriptsize $\pm.001$} & $0.014$ & $0.184${\scriptsize $\pm.001$} & $0.924${\scriptsize $\pm.002$} & $0.813${\scriptsize $\pm.029$} & $0.551${\scriptsize $\pm.006$} & $0.658${\scriptsize $\pm.002$} \\
\midrule
\textbf{LeCellModel} & $0.617${\scriptsize $\pm.017$} & $0.804${\scriptsize $\pm.003$} & $0.682${\scriptsize $\pm.008$} & $0.533$ & $1.000$ & $0.727${\scriptsize $\pm.005$} & $0.815${\scriptsize $\pm.003$} & $0.119${\scriptsize $\pm.001$} & $0.399${\scriptsize $\pm.005$} & $0.905${\scriptsize $\pm.006$} & $0.921${\scriptsize $\pm.012$} & $\mathbf{0.632}${\scriptsize $\pm.003$} & $\mathbf{0.689}${\scriptsize $\pm.003$} \\
\bottomrule
\end{tabular}%
}
\end{table}

\begin{table}[!ht]
\centering
\scriptsize
\setlength{\tabcolsep}{3pt}
\caption{Heart Cell Atlas v2 ($704{,}296$ cells): per-metric scIB results, mean $\pm$ std across $4$ seeds.}
\label{tab:scib-heart}
\resizebox{\textwidth}{!}{%
\begin{tabular}{l ccccc c ccccc c c}
\toprule
 & \multicolumn{6}{c}{\textbf{Bio-Conservation}} & \multicolumn{6}{c}{\textbf{Batch-Correction}} & \\
\cmidrule(lr){2-7}\cmidrule(lr){8-13}
Method & Iso.\ Lab.\ & NMI & ARI & Sil.\ & cLISI & \textbf{Bio} & BRAS & iLISI & kBET & GC & PCR & \textbf{Batch} & \textbf{Total} \\
\midrule
PCA            & $0.726${\scriptsize $\pm.012$} & $0.820$ & $0.692${\scriptsize $\pm.001$} & $0.578${\scriptsize $\pm.020$} & $1.000$ & $0.763${\scriptsize $\pm.006$} & $0.552${\scriptsize $\pm.016$} & $0.021${\scriptsize $\pm.001$} & $0.090${\scriptsize $\pm.003$} & $0.907${\scriptsize $\pm.001$} & $0.000$ & $0.314${\scriptsize $\pm.002$} & $0.583${\scriptsize $\pm.003$} \\
Scanorama      & $0.642$ & $0.841$ & $0.754$ & $0.572$ & $1.000$ & $0.762$ & $0.672$ & $0.044$ & $0.117$ & $0.796$ & $0.083${\scriptsize $\pm.048$} & $0.342${\scriptsize $\pm.009$} & $0.594${\scriptsize $\pm.004$} \\
Harmony        & $0.691$ & $0.931${\scriptsize $\pm.010$} & $0.948${\scriptsize $\pm.015$} & $0.643$ & $1.000$ & $0.843${\scriptsize $\pm.005$} & $0.851$ & $0.097$ & $0.271${\scriptsize $\pm.003$} & $0.854$ & $0.696${\scriptsize $\pm.035$} & $0.554${\scriptsize $\pm.007$} & $0.727${\scriptsize $\pm.004$} \\
scVI           & $0.672${\scriptsize $\pm.004$} & $0.938${\scriptsize $\pm.001$} & $0.956$ & $0.524${\scriptsize $\pm.001$} & $1.000$ & $0.818${\scriptsize $\pm.001$} & $0.835${\scriptsize $\pm.001$} & $0.066${\scriptsize $\pm.001$} & $0.114${\scriptsize $\pm.001$} & $0.900${\scriptsize $\pm.001$} & $0.801${\scriptsize $\pm.025$} & $0.543${\scriptsize $\pm.005$} & $0.708${\scriptsize $\pm.002$} \\
\midrule
\textbf{LeCellModel} & $0.835${\scriptsize $\pm.004$} & $0.935${\scriptsize $\pm.005$} & $0.952${\scriptsize $\pm.007$} & $0.524${\scriptsize $\pm.001$} & $1.000$ & $\mathbf{0.849}${\scriptsize $\pm.002$} & $0.777${\scriptsize $\pm.003$} & $0.155${\scriptsize $\pm.001$} & $0.330${\scriptsize $\pm.006$} & $0.919${\scriptsize $\pm.007$} & $0.953${\scriptsize $\pm.005$} & $\mathbf{0.627}${\scriptsize $\pm.001$} & $\mathbf{0.760}${\scriptsize $\pm.002$} \\
\bottomrule
\end{tabular}%
}
\end{table}

\begin{table}[!ht]
\centering
\scriptsize
\setlength{\tabcolsep}{3pt}
\caption{Pancreas ($16{,}382$ cells): per-metric scIB results, mean $\pm$ std across $4$ seeds.}
\label{tab:scib-pancreas}
\resizebox{\textwidth}{!}{%
\begin{tabular}{l ccccc c ccccc c c}
\toprule
 & \multicolumn{6}{c}{\textbf{Bio-Conservation}} & \multicolumn{6}{c}{\textbf{Batch-Correction}} & \\
\cmidrule(lr){2-7}\cmidrule(lr){8-13}
Method & Iso.\ Lab.\ & NMI & ARI & Sil.\ & cLISI & \textbf{Bio} & BRAS & iLISI & kBET & GC & PCR & \textbf{Batch} & \textbf{Total} \\
\midrule
PCA            & $0.656$ & $0.653$ & $0.355$ & $0.528$ & $1.000$ & $0.638$ & $0.461$ & $0.004$ & $0.114$ & $0.689$ & $0.000$ & $0.254$ & $0.485$ \\
Scanorama      & $0.524${\scriptsize $\pm.002$} & $0.860${\scriptsize $\pm.010$} & $0.904${\scriptsize $\pm.010$} & $0.559${\scriptsize $\pm.001$} & $1.000$ & $0.770${\scriptsize $\pm.004$} & $0.780${\scriptsize $\pm.004$} & $0.126$ & $0.367${\scriptsize $\pm.005$} & $0.799${\scriptsize $\pm.022$} & $0.711${\scriptsize $\pm.003$} & $0.556${\scriptsize $\pm.006$} & $0.684${\scriptsize $\pm.004$} \\
Harmony        & $0.666$ & $0.768${\scriptsize $\pm.002$} & $0.617${\scriptsize $\pm.003$} & $0.582$ & $1.000$ & $0.727${\scriptsize $\pm.001$} & $0.714$ & $0.203$ & $0.457$ & $0.762$ & $0.800$ & $0.587$ & $0.671${\scriptsize $\pm.001$} \\
scVI           & $0.719${\scriptsize $\pm.010$} & $0.918${\scriptsize $\pm.003$} & $0.952${\scriptsize $\pm.002$} & $0.568${\scriptsize $\pm.002$} & $1.000$ & $\mathbf{0.831}${\scriptsize $\pm.003$} & $0.782${\scriptsize $\pm.001$} & $0.131${\scriptsize $\pm.001$} & $0.347${\scriptsize $\pm.030$} & $0.879${\scriptsize $\pm.009$} & $0.829${\scriptsize $\pm.005$} & $0.594${\scriptsize $\pm.007$} & $0.736${\scriptsize $\pm.003$} \\
\midrule
\textbf{LeCellModel} & $0.642${\scriptsize $\pm.051$} & $0.908${\scriptsize $\pm.004$} & $0.940${\scriptsize $\pm.003$} & $0.547${\scriptsize $\pm.001$} & $1.000$ & $0.807${\scriptsize $\pm.010$} & $0.714${\scriptsize $\pm.007$} & $0.232${\scriptsize $\pm.006$} & $0.416${\scriptsize $\pm.017$} & $0.916${\scriptsize $\pm.008$} & $0.923${\scriptsize $\pm.005$} & $\mathbf{0.640}${\scriptsize $\pm.007$} & $\mathbf{0.741}${\scriptsize $\pm.008$} \\
\bottomrule
\end{tabular}%
}
\end{table}

\section{Full rare-state detection results}
\label{app:contam-full}

Full AUROC, AUPRC, and Spearman~$\rho$ (within stimulated cells, against the pathway signature z-scored on resting cells) for all four (cytokine, cell type) pairs across all four minority fractions (seed 42, one run per pair).

\begin{table}[!ht]
\centering
\scriptsize
\setlength{\tabcolsep}{3pt}
\caption{Full AUROC by (pair, $f$, method). Bold marks the best entry per row.}
\label{tab:contam-full}
\resizebox{\textwidth}{!}{%
\begin{tabular}{ll cccccccc}
\toprule
Pair & $f$ & \textbf{LeCellModel} & PCA-kNN & PCA-Mahal & PCA-IForest & scVI-IForest & scVI-recon & scVI-Mahal & scVI-kNN \\
\midrule
\multirow{4}{*}{IL-10 $\times$ CD16}
  & 0.01 & 0.835 & \textbf{0.863} & 0.820 & 0.817 & 0.847 & 0.604 & 0.819 & 0.839 \\
  & 0.03 & \textbf{0.844} & 0.822 & 0.773 & 0.738 & 0.797 & 0.561 & 0.748 & 0.769 \\
  & 0.05 & \textbf{0.854} & 0.801 & 0.745 & 0.742 & 0.704 & 0.565 & 0.719 & 0.739 \\
  & 0.10 & \textbf{0.832} & 0.751 & 0.698 & 0.704 & 0.663 & 0.557 & 0.656 & 0.683 \\
\midrule
\multirow{4}{*}{IL-10 $\times$ CD14}
  & 0.01 & \textbf{0.749} & 0.720 & 0.680 & 0.661 & 0.737 & 0.562 & 0.729 & 0.711 \\
  & 0.03 & \textbf{0.756} & 0.702 & 0.674 & 0.679 & 0.677 & 0.556 & 0.694 & 0.666 \\
  & 0.05 & \textbf{0.761} & 0.685 & 0.662 & 0.635 & 0.673 & 0.552 & 0.654 & 0.625 \\
  & 0.10 & \textbf{0.744} & 0.642 & 0.627 & 0.600 & 0.604 & 0.551 & 0.593 & 0.563 \\
\midrule
\multirow{4}{*}{IL-4 $\times$ CD14}
  & 0.01 & 0.906 & 0.911 & 0.905 & \textbf{0.912} & 0.856 & 0.548 & 0.910 & 0.866 \\
  & 0.03 & \textbf{0.885} & 0.874 & 0.852 & 0.843 & 0.835 & 0.528 & 0.834 & 0.762 \\
  & 0.05 & \textbf{0.871} & 0.848 & 0.821 & 0.811 & 0.752 & 0.528 & 0.785 & 0.730 \\
  & 0.10 & \textbf{0.845} & 0.798 & 0.769 & 0.751 & 0.697 & 0.521 & 0.701 & 0.659 \\
\midrule
\multirow{4}{*}{IL-4 $\times$ CD16 (weak)}
  & 0.01 & 0.667 & 0.724 & 0.721 & 0.720 & 0.714 & 0.467 & 0.765 & \textbf{0.775} \\
  & 0.03 & 0.686 & \textbf{0.729} & 0.702 & 0.694 & 0.675 & 0.467 & 0.713 & 0.682 \\
  & 0.05 & \textbf{0.705} & 0.701 & 0.661 & 0.658 & 0.646 & 0.464 & 0.680 & 0.667 \\
  & 0.10 & \textbf{0.690} & 0.667 & 0.618 & 0.606 & 0.619 & 0.465 & 0.608 & 0.626 \\
\bottomrule
\end{tabular}%
}
\end{table}

\begin{table}[!ht]
\centering
\scriptsize
\setlength{\tabcolsep}{3pt}
\caption{Full AUPRC by (pair, $f$, method). Bold marks the best entry per row.}
\label{tab:contam-auprc}
\resizebox{\textwidth}{!}{%
\begin{tabular}{ll cccccccc}
\toprule
Pair & $f$ & \textbf{LeCellModel} & PCA-kNN & PCA-Mahal & PCA-IForest & scVI-IForest & scVI-recon & scVI-Mahal & scVI-kNN \\
\midrule
\multirow{4}{*}{IL-10 $\times$ CD16}
  & 0.01 & 0.085 & 0.054 & 0.045 & 0.046 & \textbf{0.129} & 0.015 & 0.081 & 0.093 \\
  & 0.03 & \textbf{0.217} & 0.103 & 0.090 & 0.076 & 0.134 & 0.038 & 0.096 & 0.102 \\
  & 0.05 & \textbf{0.326} & 0.141 & 0.121 & 0.127 & 0.112 & 0.062 & 0.127 & 0.134 \\
  & 0.10 & \textbf{0.416} & 0.208 & 0.182 & 0.193 & 0.180 & 0.119 & 0.174 & 0.185 \\
\midrule
\multirow{4}{*}{IL-10 $\times$ CD14}
  & 0.01 & 0.024 & 0.020 & 0.017 & 0.017 & \textbf{0.028} & 0.012 & 0.026 & 0.024 \\
  & 0.03 & \textbf{0.071} & 0.052 & 0.049 & 0.052 & 0.055 & 0.036 & 0.059 & 0.055 \\
  & 0.05 & \textbf{0.120} & 0.080 & 0.077 & 0.069 & 0.088 & 0.059 & 0.080 & 0.076 \\
  & 0.10 & \textbf{0.203} & 0.137 & 0.134 & 0.123 & 0.132 & 0.118 & 0.127 & 0.121 \\
\midrule
\multirow{4}{*}{IL-4 $\times$ CD14}
  & 0.01 & \textbf{0.164} & 0.123 & 0.142 & 0.128 & 0.073 & 0.014 & 0.117 & 0.063 \\
  & 0.03 & \textbf{0.233} & 0.177 & 0.174 & 0.165 & 0.125 & 0.036 & 0.122 & 0.081 \\
  & 0.05 & \textbf{0.292} & 0.221 & 0.217 & 0.218 & 0.127 & 0.059 & 0.146 & 0.109 \\
  & 0.10 & \textbf{0.407} & 0.298 & 0.294 & 0.273 & 0.187 & 0.114 & 0.189 & 0.159 \\
\midrule
\multirow{4}{*}{IL-4 $\times$ CD16 (weak)}
  & 0.01 & 0.017 & 0.025 & 0.029 & 0.029 & 0.027 & 0.009 & \textbf{0.045} & 0.038 \\
  & 0.03 & 0.058 & 0.067 & \textbf{0.073} & 0.071 & 0.059 & 0.026 & \textbf{0.073} & 0.058 \\
  & 0.05 & \textbf{0.106} & 0.092 & 0.089 & 0.084 & 0.087 & 0.043 & 0.095 & 0.087 \\
  & 0.10 & \textbf{0.190} & 0.151 & 0.138 & 0.134 & 0.144 & 0.087 & 0.138 & 0.144 \\
\bottomrule
\end{tabular}%
}
\end{table}

\begin{table}[!ht]
\centering
\scriptsize
\setlength{\tabcolsep}{3pt}
\caption{Spearman~$\rho$ between per-cell score and pathway signature, computed within stimulated cells only. Bold marks the best entry per row.}
\label{tab:contam-spearman}
\resizebox{\textwidth}{!}{%
\begin{tabular}{ll cccccccc}
\toprule
Pair & $f$ & \textbf{LeCellModel} & PCA-kNN & PCA-Mahal & PCA-IForest & scVI-IForest & scVI-recon & scVI-Mahal & scVI-kNN \\
\midrule
\multirow{4}{*}{IL-10 $\times$ CD16}
  & 0.01 & 0.532 & 0.514 & 0.565 & \textbf{0.611} & 0.414 & 0.252 & 0.384 & 0.390 \\
  & 0.03 & 0.463 & 0.378 & \textbf{0.470} & 0.449 & 0.279 & 0.235 & 0.275 & 0.238 \\
  & 0.05 & \textbf{0.473} & 0.280 & 0.361 & 0.339 & 0.151 & 0.239 & 0.183 & 0.126 \\
  & 0.10 & \textbf{0.428} & 0.182 & 0.255 & 0.232 & 0.083 & 0.249 & 0.071 & 0.032 \\
\midrule
\multirow{4}{*}{IL-10 $\times$ CD14}
  & 0.01 & \textbf{0.363} & 0.172 & 0.250 & 0.265 & 0.142 & 0.252 & 0.133 & 0.072 \\
  & 0.03 & \textbf{0.315} & 0.122 & 0.240 & 0.238 & 0.104 & 0.250 & 0.029 & -0.048 \\
  & 0.05 & \textbf{0.312} & 0.091 & 0.210 & 0.222 & 0.011 & 0.255 & -0.030 & -0.101 \\
  & 0.10 & 0.251 & 0.031 & 0.138 & 0.102 & -0.030 & \textbf{0.254} & -0.097 & -0.167 \\
\midrule
\multirow{4}{*}{IL-4 $\times$ CD14}
  & 0.01 & \textbf{0.688} & 0.344 & 0.659 & 0.605 & 0.359 & 0.179 & 0.410 & 0.051 \\
  & 0.03 & \textbf{0.535} & 0.281 & 0.519 & 0.484 & 0.270 & 0.192 & 0.218 & -0.099 \\
  & 0.05 & \textbf{0.473} & 0.232 & 0.430 & 0.435 & 0.156 & 0.180 & 0.151 & -0.104 \\
  & 0.10 & \textbf{0.414} & 0.186 & 0.330 & 0.315 & 0.056 & 0.163 & 0.058 & -0.139 \\
\midrule
\multirow{4}{*}{IL-4 $\times$ CD16 (weak)}
  & 0.01 & 0.468 & 0.522 & 0.524 & \textbf{0.525} & 0.367 & 0.145 & 0.365 & 0.343 \\
  & 0.03 & 0.442 & 0.515 & \textbf{0.523} & 0.481 & 0.238 & 0.206 & 0.218 & 0.078 \\
  & 0.05 & 0.464 & 0.498 & \textbf{0.516} & 0.492 & 0.144 & 0.193 & 0.178 & 0.052 \\
  & 0.10 & 0.428 & \textbf{0.463} & 0.436 & 0.429 & 0.210 & 0.185 & 0.164 & 0.102 \\
\bottomrule
\end{tabular}%
}
\end{table}

To separate the readout from the representation, we compute $k$NN density, Mahalanobis distance and isolation forest in LeCellModel's own projection, on the same cells and runs as JEPA-SCORE (Table~\ref{tab:contam-sameemb}). On the three established-response pairs, the two distance scores fall below chance at every fraction (e.g.\ AUROC $0.27$ at $f = 1\%$ on IL-10~$\times$~CD16, against $0.84$ for JEPA-SCORE): under SIGReg, rare-and-predictive states occupy a compact, low-volume region that distance-based scores read as dense. Isolation forest is competitive only on IL-4~$\times$~CD14 at low contamination ($0.89$ at $f = 1\%$) and falls to $0.75$ at $f = 10\%$, while JEPA-SCORE stays between $0.84$ and $0.91$ on that pair. On the weaker IL-4~$\times$~CD16 response, all four scores are closer to chance.

\begin{table}[!ht]
\centering
\small
\setlength{\tabcolsep}{4pt}
\caption{Rare-state detection with simple scores computed in LeCellModel's projection: AUROC by pair and contamination fraction $f$, mean over seeds 1--3 (s.d.\ $\le 0.07$). Bold marks the best entry per row.}
\label{tab:contam-sameemb}
\begin{tabular}{ll cccc}
\toprule
Pair & $f$ & JEPA-SCORE & $k$NN & Mahal. & IForest \\
\midrule
\multirow{4}{*}{IL-10 $\times$ CD14}
  & 0.01 & \textbf{0.746} & 0.312 & 0.314 & 0.525 \\
  & 0.03 & \textbf{0.758} & 0.277 & 0.288 & 0.549 \\
  & 0.05 & \textbf{0.761} & 0.272 & 0.289 & 0.601 \\
  & 0.10 & \textbf{0.744} & 0.280 & 0.293 & 0.569 \\
\midrule
\multirow{4}{*}{IL-10 $\times$ CD16}
  & 0.01 & \textbf{0.837} & 0.273 & 0.273 & 0.576 \\
  & 0.03 & \textbf{0.865} & 0.184 & 0.223 & 0.617 \\
  & 0.05 & \textbf{0.864} & 0.178 & 0.219 & 0.639 \\
  & 0.10 & \textbf{0.840} & 0.203 & 0.250 & 0.625 \\
\midrule
\multirow{4}{*}{IL-4 $\times$ CD14}
  & 0.01 & \textbf{0.907} & 0.140 & 0.290 & 0.893 \\
  & 0.03 & \textbf{0.882} & 0.153 & 0.249 & 0.863 \\
  & 0.05 & \textbf{0.866} & 0.164 & 0.264 & 0.849 \\
  & 0.10 & \textbf{0.844} & 0.185 & 0.288 & 0.753 \\
\midrule
\multirow{4}{*}{IL-4 $\times$ CD16 (weak)}
  & 0.01 & \textbf{0.630} & 0.604 & 0.577 & 0.495 \\
  & 0.03 & \textbf{0.675} & 0.510 & 0.509 & 0.497 \\
  & 0.05 & \textbf{0.701} & 0.412 & 0.447 & 0.499 \\
  & 0.10 & \textbf{0.688} & 0.383 & 0.421 & 0.506 \\
\bottomrule
\end{tabular}
\end{table}

\FloatBarrier
\section{Cytokine-stimulation panel: pair details}
\label{app:contam-pairs}

All pairs draw from a fixed pool of $16{,}710$ resting CD16 monocytes and $110{,}999$ resting CD14 monocytes. We consider four cytokine--cell-type settings: \emph{IL-4~$\times$~CD14 Mono}, a canonical M2-like IL-4/STAT6 response; \emph{IL-10~$\times$~CD14 Mono}, a canonical anti-inflammatory IL-10/STAT3 response; \emph{IL-10~$\times$~CD16 Mono}, an IL-10/STAT3 response in a distinct CD16$^{+}$ background; and \emph{IL-4~$\times$~CD16 Mono}, a weaker IL-4-response setting reported in Appendix~\ref{app:contam-full}. Density-based baselines use scVI with \texttt{n\_latent=30}, 100 epochs.

\paragraph{Pathway signatures.} The Spearman~$\rho$ in Table~\ref{tab:contam-spearman} uses the mean z-score (z-scored on resting cells) of an up-set minus a down-set: \textbf{IL-10}, up = \{CD163, CD163L1, IRAK3, MARCO, MRC1, F13A1\} (STAT3 targets), down = \{IL1B, IL6, CXCL10, CCL3, CCL4, CXCL8\} (IL-10-suppressed pro-inflammatory mediators); \textbf{IL-4}, up = \{CCL17, CCL22, MRC1, ALOX15, CCL18, CHI3L1, F13A1\} (STAT6 targets), no down-set.

\section{Rare-program discovery: statistics and cluster selection}
\label{app:gradclust}

\paragraph{Bivariate Moran's I on the embedding graph.}
For each gene $g$, bivariate Moran's $I_g$ measures whether gene expression and JEPA score are spatially anti-correlated on the embedding graph: $I_g < 0$ means $g$ is elevated in cells whose neighbors have low JEPA score, marking it as a candidate gene of the rare subpopulation. $W$ is the Gaussian-kernel-weighted $k$NN graph ($k=15$, row-stochastic) on the LeCellModel projection; significance uses the analytical $z$-score from Hotspot~\citep{hotspot}, with BH-FDR $<0.05$.

Before computing $I_g$ we apply a logistic transform to the raw score, centering at the 5th-percentile score $c = \hat{F}_{5\%}(S)$, scaling by the score standard deviation $\sigma_S$, and squashing with a sigmoid of steepness $\beta=2$:
\[
  \tilde{S}(x) \;=\; \sigma\!\left(\beta\,\frac{S(x) - c}{\sigma_S}\right).
\]
This maps the rare anomalous tail ($S \ll c$) toward 0 and the bulk ($S \gg c$) toward 1.
JEPA-coupled genes ($I_g < 0$, BH-FDR $< 0.05$) are clustered into modules by their per-cell coupling profile $\mathbf{q}_g \in \mathbb{R}^N$, where $(\mathbf{q}_g)_n = \tilde{x}_{g,n}\cdot(W\tilde{S})_n$ is the product of gene $g$'s standardized expression and the neighborhood-smoothed score at cell $n$. Genes in the same module are elevated in the same low-score cells, not merely co-expressed globally. Module~1 (ranked by $|I|\cdot\sqrt{n_{\mathrm{genes}}}$) is the recovered program reported in Section~\ref{sec:rare} (Table~\ref{tab:morans-program}).

\begin{table}[!ht]
\centering
\small
\setlength{\tabcolsep}{6pt}
\caption{Top-15 members of the recovered program (Module 1) per cohort, ranked by $|I_g|$ within the module. $I_g$ is the raw bivariate Moran's I (range $[-1, 1]$); negative values indicate the gene is elevated in low-JEPA-score (rare) cells. Module 1 is the dominant co-coupling cluster among JEPA-coupled genes (BH-FDR $< 0.05$, $I_g < 0$). Mono--mac Module 1: $n_{\text{genes}}=47$, mean $I=-0.732$. CD4 Module 1: $n_{\text{genes}}=48$, mean $I=-0.784$.}
\label{tab:morans-program}
\begin{tabular}{rlr@{\hskip 24pt}rlr}
\toprule
\multicolumn{3}{c}{\textbf{Monocyte-derived macrophages (Module 1)}} & \multicolumn{3}{c}{\textbf{CD4 T cells (Module 1)}} \\
\cmidrule(lr){1-3}\cmidrule(lr){4-6}
Rank & Gene & $I_g$ & Rank & Gene & $I_g$ \\
\midrule
1  & CHI3L1   & $-0.459$ & 1  & FOXP3     & $-0.484$ \\
2  & ADAMDEC1 & $-0.429$ & 2  & TNFRSF18  & $-0.417$ \\
3  & MMP9     & $-0.415$ & 3  & IL2RA     & $-0.416$ \\
4  & CHIT1    & $-0.355$ & 4  & TNFRSF9   & $-0.408$ \\
5  & CTSK     & $-0.353$ & 5  & TNFRSF4   & $-0.398$ \\
6  & MREG     & $-0.327$ & 6  & TIGIT     & $-0.397$ \\
7  & RARRES1  & $-0.325$ & 7  & CD177     & $-0.397$ \\
8  & A2M      & $-0.325$ & 8  & LAYN      & $-0.385$ \\
9  & SIGLEC15 & $-0.319$ & 9  & CTLA4     & $-0.370$ \\
10 & GPC4     & $-0.317$ & 10 & LINC02099 & $-0.311$ \\
11 & LILRB4   & $-0.315$ & 11 & TBC1D4    & $-0.303$ \\
12 & MMP7     & $-0.315$ & 12 & IL1R2     & $-0.297$ \\
13 & SPP1     & $-0.297$ & 13 & ENTPD1    & $-0.297$ \\
14 & PALLD    & $-0.295$ & 14 & CCR8      & $-0.296$ \\
15 & FCMR     & $-0.294$ & 15 & IKZF2     & $-0.281$ \\
\bottomrule
\end{tabular}
\end{table}

\paragraph{Cluster selection for gradient attribution.}
Gradient attribution requires a concrete cell cluster. We identify connected neighborhoods of low-JEPA-score cells via a pit-finding routine; the exact cluster boundary does not materially affect the attributed genes. The routine operates on a cosine kNN graph ($k=10$) restricted to the bottom $10\%$ of cells by JEPA-score (the \emph{subgraph}), with a data-adaptive distance threshold $\delta$ --- the $10^{\text{th}}$ percentile of kNN edge distances within the subgraph --- controlling which neighbors are traversable.
\begin{enumerate}
\item Smooth the JEPA-score: replace each cell's score with the mean of itself and neighbors at distance $\leq\delta$, one pass.
\item From seeds at the bottom $0.5\%$ of JEPA-score (all cells), follow the lowest-scoring neighbor at distance $\leq\delta$ until no lower neighbor exists; the landing cell is a local minimum (``pit bottom'').
\item Group seeds by pit bottom; discard pits with fewer than $\max(3,\, 5\% \text{ of seeds})$ converging seeds.
\item Iteratively merge the least-prominent adjacent pit pair: the prominence of pit $B$ is the lowest cross-boundary score between $A$ and $B$ minus the score of $B$'s bottom (i.e.\ how much the score rises to leave $B$); merge until all remaining pits are separated by more than $0.3\,\sigma_S$, where $\sigma_S$ is the score standard deviation within the subgraph.
\item Expand each surviving pit into a core cluster by BFS within distance $\delta$.
\end{enumerate}
For each core cluster $C$ we compute the input$\times$gradient attribution $x \odot \nabla_x S(x)$ at every cell via automatic differentiation, then rank genes by the absolute cluster mean, $\lvert \mathbb{E}_{x\in C}[x \odot \nabla_x S] \rvert$. Gene identity is not used in cluster selection; the procedure sees only the JEPA-score and the embedding geometry. All case-study results in Section~\ref{sec:attr-case} use these defaults.

\section{Predictive selectivity: computation details}
\label{app:sel}

For each cell $x$ with full-cell projection $p = F(x)$ (log1p-CP10k input), $\mathrm{cos}_1 = \max_g \cos(F(x^{(g)}), p)$ where $x^{(g)}$ retains only gene $g$; $\mathrm{cos}_2$ further adds the best second gene on top of $g^*$, both exhaustive over all $G$ genes, reported as population medians.  The CD177\textsuperscript{+} Treg cluster uses all $n=182$ cells (FOXP3$>1$ \& CD177$>1$); the CD4 T-cell pool and profibrotic monocyte-derived macrophage groups are each capped at $n=500$ (uniform subsample, seed 42).  The scVI baseline applies the same procedure to the posterior mean $\mu_z$ with raw-count inputs; scVI is trained with \texttt{n\_latent}$=40$, \texttt{n\_layers}$=2$, \texttt{n\_hidden}$=256$, donor batch key, up to 200 epochs with early stopping.

\section{Stability of JEPA-score rankings}
\label{app:stability}
We retrain each IPF cohort with four seeds (42, 1, 2, 3) and, at seed 42, under two further augmentation settings (global/local mask rates $0.2/0.6$ and $0.1/0.4$, against $0.5/0.7$ in Table~\ref{tab:lejac-hparams-ipf}) and with Poisson resampling disabled. Runs score the same cells in the same order, so rankings are compared cell by cell (Table~\ref{tab:stability}).
Across seeds, per-cell scores correlate at mean pairwise Spearman $\rho = 0.965$ (monocyte--macrophage, $257{,}893$ cells) and $0.843$ (lymphoid, $159{,}652$ cells). About $90\%$ of each run's rarest 1\% is shared with any other run. The cells that differ sit just past the cutoff: their median score percentile in the other run is $1.13$--$1.14\%$, and $99.9\%$ (monocyte) and $99.6\%$ (lymphoid) fall within its rarest 2\%. Within the shared tail, the ordering is preserved (Spearman $0.82$--$0.89$), and a consensus of three runs recovers $91$--$92\%$ of the held-out run's rare set. The rare tail is nearly as stable across augmentation settings: of the cells whose rare-1\% membership changes between mask regimes, $90.6\%$ (monocyte) and $94.2\%$ (lymphoid) stay within the rarest 2\%, and disabling Poisson resampling gives the same picture.

\begin{table}[!ht]
\centering
\small
\setlength{\tabcolsep}{4pt}
\caption{Stability of per-cell JEPA-scores in the IPF cohorts. $\rho$: mean pairwise Spearman correlation of per-cell scores; overlap: mean fraction of one run's rarest 1\% that is also in the other run's; $\le$2\%: fraction of cells whose rare-1\% membership changes that fall within the other run's rarest 2\%.}
\label{tab:stability}
\begin{tabular}{l ccc ccc}
\toprule
 & \multicolumn{3}{c}{\textbf{Monocyte--macrophage}} & \multicolumn{3}{c}{\textbf{Lymphoid}} \\
\cmidrule(lr){2-4}\cmidrule(lr){5-7}
Comparison & $\rho$ & Overlap & $\le$2\% & $\rho$ & Overlap & $\le$2\% \\
\midrule
Seeds 42, 1, 2, 3                         & 0.965 & 90.2\% & 99.9\% & 0.843 & 89.5\% & 99.6\% \\
Mask rates $0.1/0.4$, $0.2/0.6$, $0.5/0.7$ & 0.936 & 80.6\% & 90.6\% & 0.877 & 88.2\% & 94.2\% \\
Poisson resampling off vs.\ on            & 0.954 & 87.1\% & 95.0\% & 0.894 & 93.1\% & 99.5\% \\
\bottomrule
\end{tabular}
\end{table}

\section{Gradient attribution against pathway targets and differential expression}
\label{app:attr-de}
 We rank genes by the absolute cluster-mean input$\times$gradient, $\lvert \mathbb{E}_{x\in C}[x \odot \nabla_x S] \rvert$, and compare with the raw gradient, $\lvert \mathbb{E}_{x\in
C}[\nabla_x S] \rvert$, and with Wilcoxon differential expression (Table~\ref{tab:attr-de}). Without any pathway list, the gradient ranks canonical STAT3 and STAT6 targets at the top on
the CD14 pairs (F13A1 and ALOX15 first), and lower on the CD16 pairs. Where the two rankings disagree, the gradient favors genes expressed in a subset of the cluster (CD177, SIGLEC15,
MARCO, CCL18), while DE favors broadly expressed markers such as SPP1. The gradient thus captures local drivers of the score rather than abundance differences between groups, and needs
no predefined contrast group.

\begin{table}[!ht]
\centering
\small
\setlength{\tabcolsep}{4pt}
\caption{Gene ranks by input$\times$gradient, raw gradient and differential expression (Wilcoxon rank-sum, ranked by test statistic). IPF: cluster against the rest of its cell type, ranks over $3{,}000$ genes. Cytokine panel ($f = 5\%$): stimulated against control cells, ranks over $2{,}000$ genes. Expressed: share of the cluster's (IPF) or stimulated (cytokine) cells with non-zero expression.}
\label{tab:attr-de}
\begin{tabular}{ll ccc c}
\toprule
Gene & Setting & Input$\times$grad. & Raw grad. & DE & Expressed \\
\midrule
CD177    & IPF, CD4 T cells               & 2 & 1 & 25 & 37\% \\
SIGLEC15 & IPF, mono.-derived macrophages & 9 & 4 & 43 & 50\% \\
SPP1     & IPF, mono.-derived macrophages & 414 & 2{,}875 & 8 & 92\% \\
\midrule
F13A1  & IL-10 $\times$ CD14 (seed 1)     & 1 & 12 & 1 & 59\% \\
ALOX15 & IL-4 $\times$ CD14 (seed 1)      & 1 & 3 & 2 & 68\% \\
CCL18  & IL-4 $\times$ CD14 (seed 1)      & 8 & 25 & 93 & 14\% \\
MARCO  & IL-10 $\times$ CD16 (seeds 1--3) & 3--7 & 1 & 26--35 & 28--31\% \\
CD163  & IL-10 $\times$ CD16 (seeds 1--3) & 4--19 & 14--65 & 5--7 & 69--72\% \\
F13A1  & IL-10 $\times$ CD16 (seeds 1--3) & 74--91 & 27--43 & 84--91 & 16--18\% \\
ALOX15 & IL-4 $\times$ CD16 (seeds 1--3)  & 8--19 & 1--2 & 43 & 14--15\% \\
\bottomrule
\end{tabular}
\end{table}

\section{Negative control on healthy heart tissue}
\label{app:negative-control}

As a negative control we use healthy tissue, where no disease program is present: the subsampled Heart Cell Atlas distributed with scvi-tools~\citep{heart-ca}. We train with this dataset's own configuration: batch size $512$, $\lambda_{\mathrm{sig}} = 0.4$, adversary weight $0.8$ with progress factor $2.0$, one global and four local views with mask rates $0.2/0.5$, $d = k = 256$, no displacement auxiliary, $150$ epochs. In this model, the lowest-scoring regions include an activated venous-endothelial state ($n = 46$; top input$\times$gradient genes ACKR1, SELE and VCAM1). We then remove every cell at or below the 20th percentile of JEPA-score within its cell type ($18{,}641 \to 14{,}908$ cells) and retrain on the remaining cells. The retrained model finds one low-score region instead of four; the remaining ventricular-cardiomyocyte region lies $0.67$ standard deviations below the mean, against $1.71$ before pruning, and the score distribution tightens (mean $-818 \to -747$, s.d.\ $88 \to 79$): homogeneous populations receive more uniform scores.


\end{document}